\documentclass[]{mosi}
\usepackage{placeins}

\newcommand{\faPlayCircle}{\raisebox{-0.2ex}{\includegraphics[height=2.0ex]{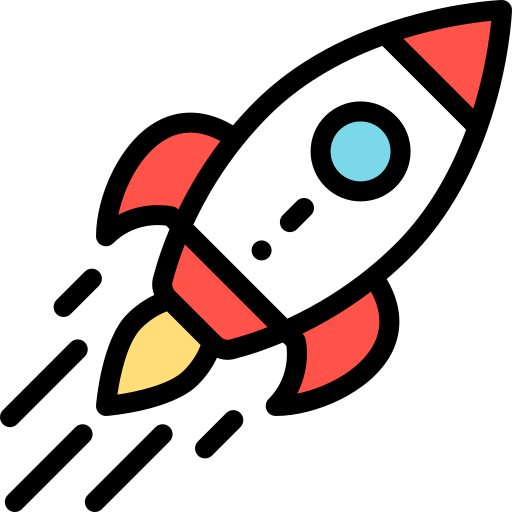}}}
\newcommand{\faCogs}{\raisebox{-0.2ex}{\includegraphics[height=2.0ex]{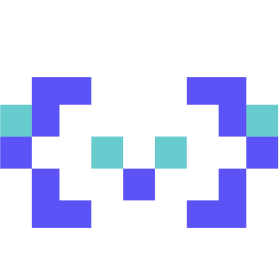}}}
\newcommand{\hflogo}{\raisebox{-0.2ex}{\includegraphics[height=2.0ex]{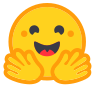}}}

\title{MOSS-Audio Technical Report}

\author{OpenMOSS Team}

\abstract{
MOSS-Audio is a unified audio-language model  for speech, environmental sound, and music understanding, supporting audio captioning, time-aware question answering, timestamped transcription, and audio-grounded reasoning. MOSS-Audio couples a dedicated audio encoder with a modality adapter and a large language model: the encoder produces 12.5 Hz temporal representations, the adapter projects them into the decoder space, and the decoder generates autoregressive text outputs. Two design choices are central to the system: \textbf{DeepStack cross-layer feature injection}, which exposes the decoder to acoustic information from multiple encoder depths, and \textbf{time markers}, which provide explicit temporal cues by inserting timestamp markers into the audio-token stream. At the data level, we design an event-preserving audio annotation pipeline that segments raw audio at coherent event boundaries, applies branch-specific annotation to speech, music, and general audio, and merges the results into unified captions for pretraining. The intermediate branch-specific captions are further retained to support the construction of task-oriented SFT data. The model is pretrained on large-scale audio-language data, with time-aware objectives incorporated to support temporal grounding, and then undergoes multi-stage post-training to enhance instruction following and audio-grounded reasoning. We release 4B and 8B variants in both Instruct and Thinking configurations. MOSS-Audio achieves strong performance across general audio understanding, speech captioning, ASR, and timestamped ASR, positioning it as a promising understanding foundation for future voice agents.
}

\checkdata[\raisebox{-0.1ex}{\faPlayCircle}\hspace{0.4em}Demo]{\url{https://openmoss.github.io/MOSS-Audio/}}
\checkdata[\raisebox{-0.1ex}{\hflogo}\hspace{0.4em}Hugging Face]{\url{https://huggingface.co/collections/OpenMOSS-Team/moss-audio}}
\checkdata[\raisebox{-0.1ex}{\faCogs}\hspace{0.4em}ModelScope]{\url{https://modelscope.cn/collections/openmoss/MOSS-Audio}}

\begin{document}
\maketitle
\begingroup
\renewcommand{\thefootnote}{\fnsymbol{footnote}}
\setcounter{footnote}{1}
\footnotetext{Full contributors can be found in the Contributors section.}
\endgroup


\begin{figure*}[t]
  \centering
  \includegraphics[width=.9\linewidth]{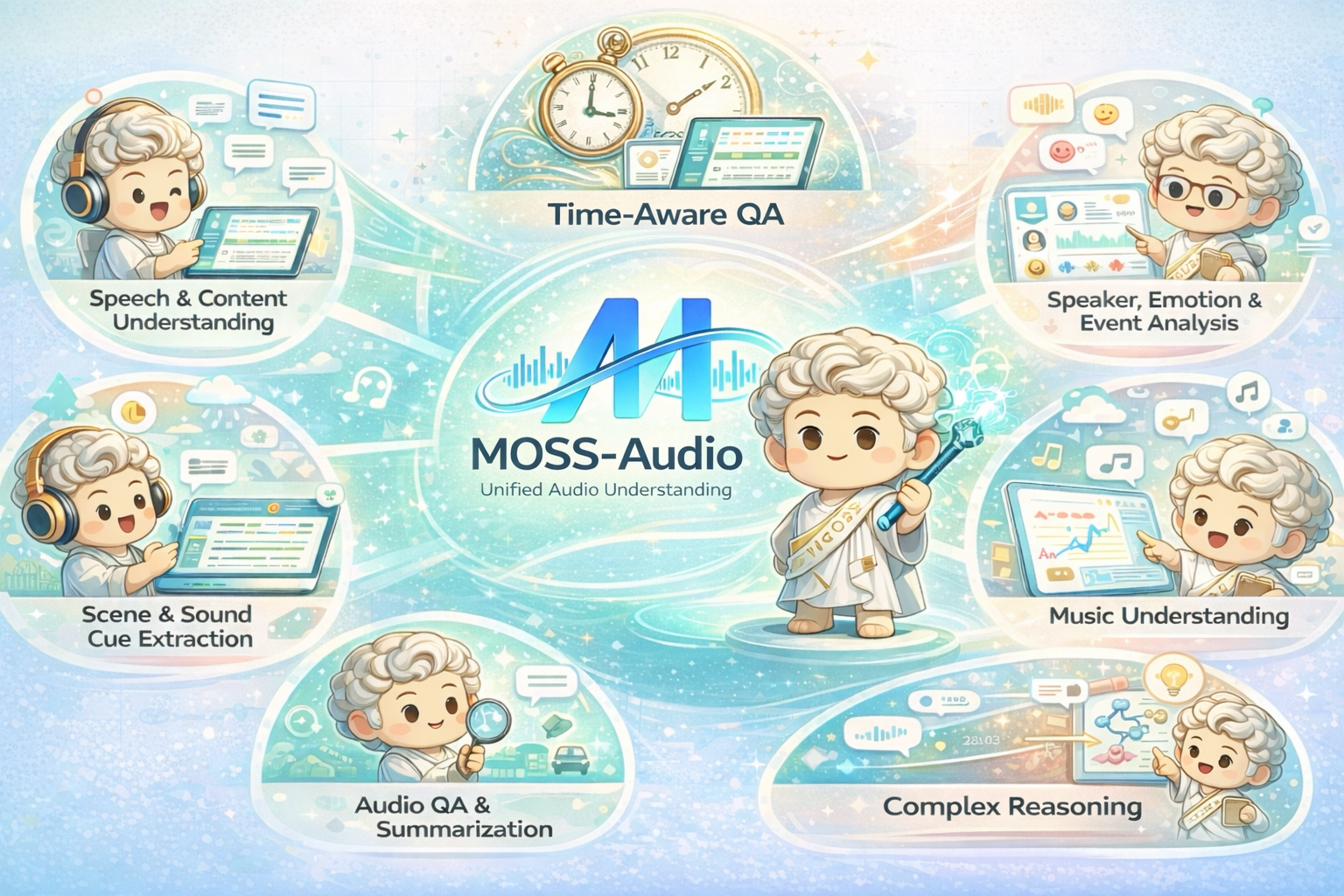}
  \caption{MOSS-Audio performs unified modeling over complex real-world audio, supporting speech understanding, environmental sound understanding, music understanding, audio captioning, time-aware QA, and complex reasoning.}
  \label{fig:moss_audio_image}
\end{figure*}

\section{Introduction}

Audio is a primary modality for perceiving language, acoustic events, environments, music, and social context. 
A speech recording contains not only words, but also speaker traits, prosody, emotion, turn-taking cues, and temporal structure \cite{wang2025mmsu}; real-world audio may further contain environmental sounds, music, overlapping events, and long-range dependencies that cannot be reduced to transcripts alone \cite{gemmeke2017audio,kim2019audiocaps,mei2024wavcaps}. 
As audio-language models move beyond automatic speech recognition \cite{radford2023robust}, a central goal is to build unified systems that can understand heterogeneous audio signals, follow natural language instructions, and produce temporally grounded textual outputs \cite{chu2023qwenaudio,tang2024salmonn,kong2024audioflamingo}.

This goal is particularly important for voice agents, where the audio model is not merely a transcription module but the perceptual and reasoning foundation for interpreting user speech, acoustic context, and time-sensitive events before downstream tools generate responses or execute actions \cite{chu2024qwen2,qwen2.5omni}. 
A capable audio understanding foundation model should therefore support multiple capabilities within one interface, including speech transcription, speech and audio captioning, music and environmental sound understanding, timestamped transcription, time-aware question answering, and audio-grounded reasoning \cite{chu2023qwenaudio,tang2024salmonn,ghosh2024gama,ghosh2026audioflamingonext}.

Building such a unified model remains challenging. 
Different tasks depend on different levels of acoustic abstraction: ASR requires fine-grained phonetic and lexical information, speech captioning relies on prosody and speaker attributes, environmental sound understanding often depends on short transient events, and reasoning-oriented tasks require semantic integration over longer contexts \cite{chu2023qwenaudio,ghosh2024gama,wang2025mmsu}. 
Moreover, many audio-language tasks are inherently temporal, requiring the model to determine not only what happens but also when it happens \cite{sridhar2025temporalAQA,sakshi2025mmau,ma2025mmar}. 
These requirements place pressure on both model architecture and data construction, making simple extensions of ASR-style training insufficient for broad audio understanding.

In this report, we present \textbf{MOSS-Audio}, a unified audio-language model family for speech, environmental sound, and music understanding. 
MOSS-Audio supports ASR, audio captioning, speech captioning, timestamped transcription, time-aware question answering, and audio-grounded reasoning within a single autoregressive text-generation framework. 
Given an audio input and a natural language instruction, the model generates task-specific textual outputs while sharing the same audio representation and language decoding interface.

MOSS-Audio follows an encoder--adapter--decoder architecture widely used in recent large audio-language models \cite{gong2023jointaudio,tang2024salmonn,chu2024qwen2}. 
A dedicated audio encoder produces compact temporal representations at 12.5 Hz, a modality adapter projects audio features into the language-model space, and a large language model generates autoregressive text conditioned on both the audio input and the instruction. 
This modular design combines an audio front-end specialized for broad acoustic understanding with the instruction-following and generation capabilities of modern language models.

Two architectural choices are central to MOSS-Audio. 
First, we introduce \textbf{DeepStack cross-layer feature injection} for audio-language modeling. 
Instead of passing only the final encoder representation to the language model, MOSS-Audio exposes the decoder to features from multiple encoder depths. 
This reduces the bottleneck of relying on a single final-layer representation and preserves acoustic evidence at different granularities, including low-level time-frequency patterns, transient events, prosodic cues, and high-level semantic information \cite{meng2024deepstack,ghosh2024gama}. 
Second, we introduce \textbf{explicit time markers} into the audio representation sequence. 
Rather than treating timestamps as external post-processing, MOSS-Audio makes temporal information part of the model context, enabling timestamped transcription and time-aware audio question answering to be learned directly through generation \cite{radford2023robust,sridhar2025temporalAQA}.

The data pipeline is also designed for unified audio understanding. 
MOSS-Audio is trained on speech, music, and general audio data, using an event-preserving segmentation strategy that avoids arbitrary fixed-window cuts when complete acoustic events should be retained. 
Segmented audio is routed into branch-specific annotation pipelines for speech, music, and general audio, and the resulting annotations are converted into unified caption and instruction formats. 
This allows heterogeneous supervision from event labels, captions, speech transcripts, and instruction data to be learned under a common language-modeling objective \cite{gemmeke2017audio,kim2019audiocaps,mei2024wavcaps,chu2023qwenaudio}.

The training pipeline integrates ASR, audio captioning, timestamp ASR, and text modeling during pre-training, followed by staged post-training for instruction following and audio-grounded reasoning. 
This produces two complementary model types. 
The \textbf{Instruct} variants are optimized for direct instruction following and stable task execution, making them suitable for transcription, captioning, and timestamp-oriented tasks. 
The \textbf{Thinking} variants are optimized for reasoning-heavy audio understanding, where the model must integrate speech, non-speech events, temporal cues, and task instructions before producing an answer. 
We release both 4B and 8B models in these two configurations: \textbf{MOSS-Audio-4B-Instruct}, \textbf{MOSS-Audio-4B-Thinking}, \textbf{MOSS-Audio-8B-Instruct}, and \textbf{MOSS-Audio-8B-Thinking}.

Empirical results show that MOSS-Audio achieves strong performance across general audio understanding, speech captioning, ASR, and timestamped ASR. 
The Thinking variants show advantages on reasoning-oriented audio understanding benchmarks, while the Instruct variants provide stronger direct task execution for transcription and captioning. 
These results indicate that a unified audio-language model can support both precise audio recognition and higher-level audio reasoning, positioning MOSS-Audio as a promising understanding foundation for future voice agents \cite{sakshi2025mmau,kumar2025mmaupro,ma2025mmar,wang2025mmsu}.

Overall, this report makes the following contributions:
\begin{itemize}
    \item We present \textbf{MOSS-Audio}, a unified audio-language model family with 4B and 8B \textbf{Instruct} and \textbf{Thinking} variants, achieving state-of-the-art performance across general audio understanding, speech captioning, ASR, and timestamped ASR.
    \item We introduce \textbf{DeepStack cross-layer feature injection} in audio-language models, which preserves multi-level acoustic evidence for language-model decoding.
    \item We incorporate \textbf{explicit time markers} to support temporally grounded generation, including timestamped transcription and time-aware audio question answering.
    \item We build a broad audio-language data pipeline based on event-preserving segmentation, branch-specific annotation, and unified caption merging, producing an annotated audio dataset at the scale of millions of hours.
\end{itemize}

\section{Architecture}

\begin{figure*}[t]
  \centering
  \includegraphics[width=0.9\linewidth]{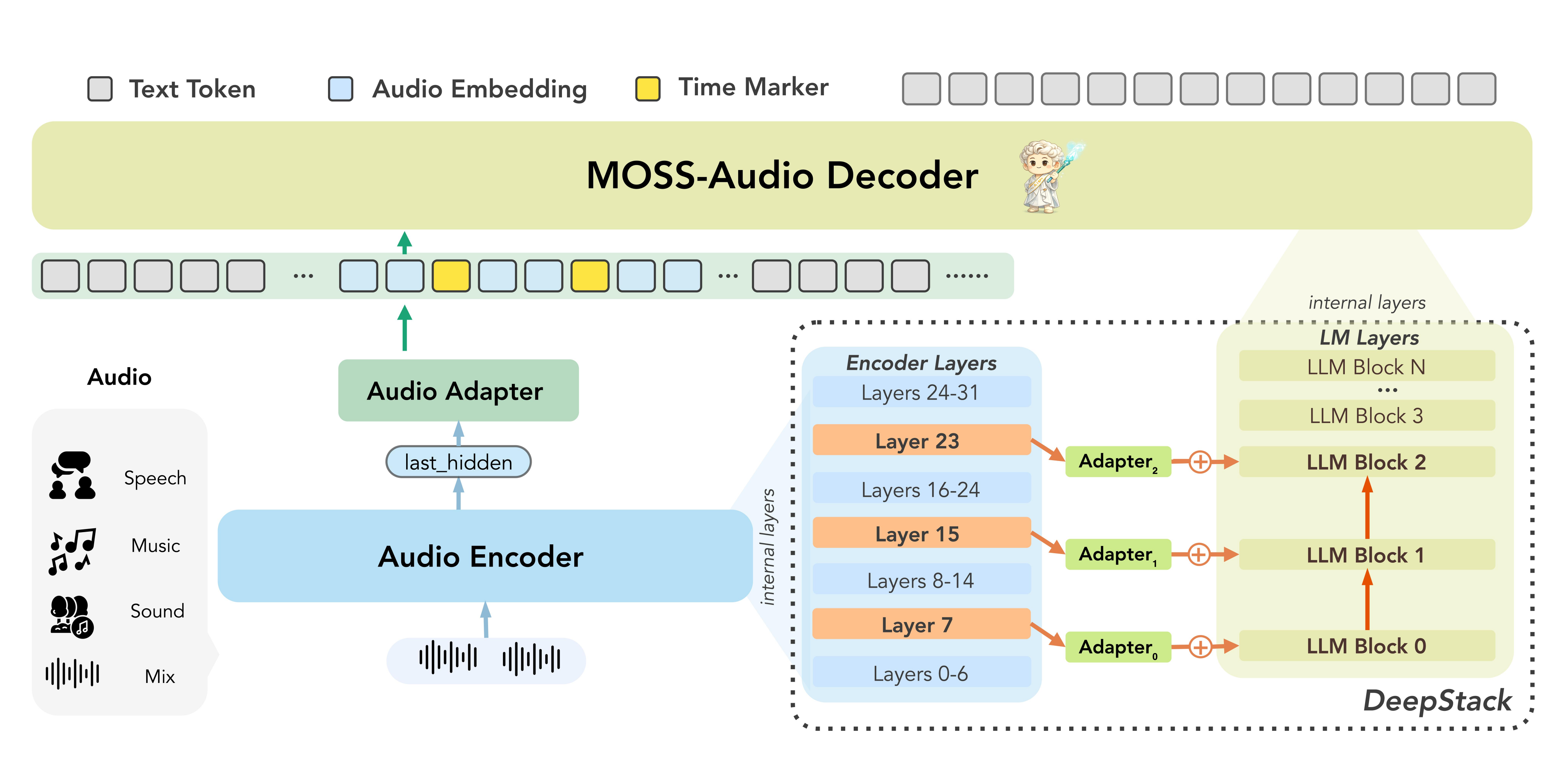}
  \caption{Architecture of MOSS-Audio.}
  \label{fig:moss_audio_architecture}
\end{figure*}

\subsection{Overview}

Most existing audio-language models reuse off-the-shelf speech encoders or frozen frontends originally trained for automatic speech recognition (ASR). Such frontends transcribe speech efficiently, but they are optimized for a single narrow objective---mapping acoustic signals to lexical tokens, and tend to discard speaker characteristics, prosodic cues, environmental context, and musical structure. Unified audio understanding instead requires a frontend that preserves a far broader set of acoustic attributes and aligns them with the semantic space of a general-purpose language model. We therefore design MOSS-Audio as an end-to-end audio-conditioned language model whose audio encoder is trained from scratch for this purpose. As shown in Figure~\ref{fig:moss_audio_architecture}, the model uses a language model as its backbone and comprises three trainable components: a dedicated audio encoder, two GatedMLP cross-modal adapters, and a decoder. Given an input waveform, the encoder converts log-mel features into a sequence of continuous temporal representations. A primary adapter projects the final encoder output into decoder's hidden space so that audio embeddings can be consumed alongside textual instructions, while a parallel DeepStack-style pathway extracts intermediate encoder states, aggregates them through a merge adapter, and injects the resulting cross-layer features into the early decoder layers; both adapters use the same GatedMLP projection. Conditioned on these representations, the decoder then performs autoregressive generation for transcription, captioning, audio question answering, temporal localization, and reasoning-oriented audio understanding.


\subsection{MOSS Audio Encoder}


To obtain a strong and robust audio encoder, we train an Audio Encoder entirely from scratch on millions of hours of diverse audio data with ASR, AST and Audio Caption tasks. The $\sim$0.6B parameter encoder module processes 128-channel log-mel spectrograms via three stride-2 Conv2D layers, achieving an 8$\times$ temporal downsampling to yield a highly efficient 12.5 Hz token rate. These features are then processed by a 32-layer Transformer backbone with a hidden dimension of 1280. To efficiently handle long-context inputs, the encoder eschews global self-attention in favor of  sliding window attention restricted to a maximum of 100 frames (8 seconds). This localized attention scales linearly with audio length, significantly reducing memory consumption and enabling real-time KV-caching, gracefully delegating long-range semantic reasoning to the language model while ensuring robust local acoustic modeling.

\subsection{DeepStack Cross-Layer Feature Injection}

Using only the final-layer output of a deep encoder tends to lose low-level acoustic details such as prosody, transient events, and local time-frequency structure. Layer-wise analyses of self-supervised speech models show that acoustic and speaker-related cues concentrate in lower and intermediate layers while deeper layers drift toward lexical and semantic content~\cite{pasad2021layerwise,chen2022wavlm}, and that a learnable combination of all layers consistently outperforms the last-layer representation across diverse speech tasks~\cite{yang2021superb}. For unified audio understanding, these fine-grained cues are essential: rhythm and timbre inform speaker and emotion analysis, transient events underlie environmental sound detection, and local spectral structure supports music understanding. A single final-layer representation therefore cannot capture the full range of granularities that downstream audio tasks require.

To preserve acoustic information across levels of abstraction, we adopt DeepStack-style cross-layer feature injection, which exposes multiple encoder depths to the language-model backbone~\cite{meng2024deepstack,bai2025qwen3vl}. Beyond the final encoder output consumed by the primary audio adapter, we extract intermediate hidden states from encoder layers and pass them through a separate merge adapter, and inject the resulting features into selected early layers of the decoder. The merge adapter uses the GatedMLP projection as the primary adapter, mapping audio features into the language-model hidden space. In this way the primary adapter supplies the main final-layer representation, while the merge adapter contributes complementary low- and mid-level acoustic evidence, giving the decoder a multi-granularity view of the audio without enlarging the encoder.

\subsection{Time-aware Modeling}

Temporal grounding is essential for audio understanding tasks such as timestamped speech recognition, acoustic event localization, and time-aware audio question answering. In the absence of explicit temporal cues, the language model must infer event timing only from the relative positions of audio tokens, which becomes increasingly unreliable for long-form audio. To expose absolute time information to the decoder, MOSS-Audio interleaves explicit elapsed-time markers into the audio-conditioned input sequence.

Following the timestamp-aware modeling strategy of MOSS Transcribe Diarize~\cite{yu2026mosstranscribediarize}, we insert numerical time markers between blocks of audio features. The audio encoder produces representations at 12.5 Hz, so 25 consecutive audio features correspond to 2 seconds. We therefore append a time marker after every 25 audio features, yielding an interleaved sequence with markers such as ''\texttt{2}``, ''\texttt{4}``, ''\texttt{6}``, and ''\texttt{8}``, where each marker indicates the elapsed time in seconds at that position. These markers are embedded and processed jointly with the adapted audio representations by the language model, providing explicit temporal anchors for timestamp generation, event localization, and time-aware audio reasoning within a unified autoregressive framework.

\section{Data Pipeline}

\begin{figure}[t]
    \centering
    \includegraphics[width=.9\textwidth]{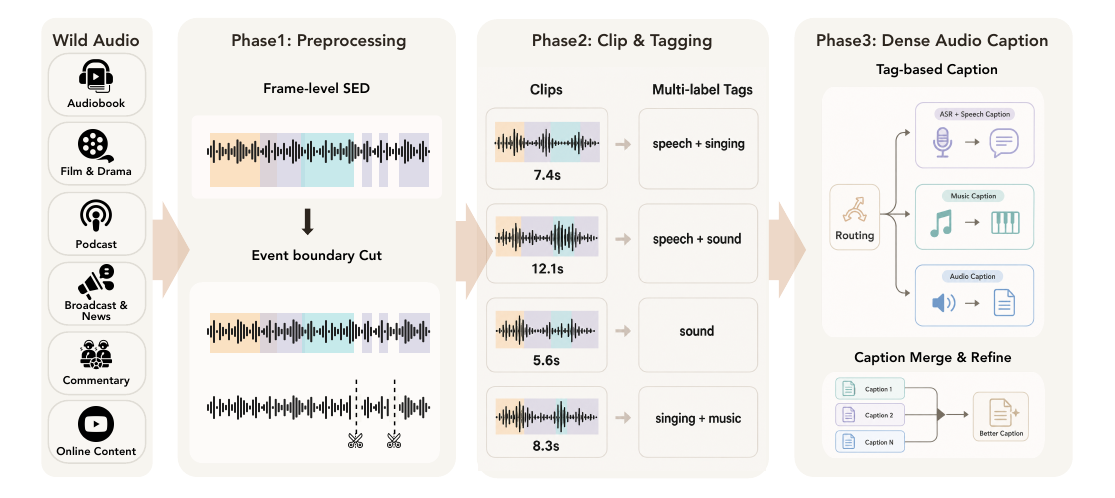}
    \caption{Overview of the data pipeline. Wild audio is segmented by event boundaries, tagged with audio labels, routed to branch-specific captioning modules, and finally merged into a unified caption for model training.}
    \label{fig:data_pipeline}
\end{figure}

MOSS-Audio uses a branched data engine for data construction. 
For wild audio, the data engine is centered on classification-guided annotation. 
Rather than applying a uniform captioning procedure to every recording, MOSS-Audio first preserves complete acoustic events through segmentation, then assigns each segment a multi-label audio profile according to its detected event composition. 
This captioning pipeline produces content-adaptive, instance-specific supervision for each audio segment, making it well suited for training audio understanding models on heterogeneous real-world audio.
\subsection{Event Segmentation}\label{sec:event_segmentation}

The pipeline begins with segmentation. Instead of cutting raw audio at fixed time intervals, we segment at natural event boundaries to produce acoustically coherent clips with intact sound events. We first run a frame-level sound event detection model on each audio file to obtain timestamped event labels under the AudioSet taxonomy, using a BEATs \cite{chen2023beats} backbone trained within the PretrainedSED framework \cite{schmid2024effectivepretrainingaudiotransformers}. We then apply a merge-and-cut procedure over the detected events. Vocal and speech-related events are merged with a gap tolerance to keep speaker turns and continuous utterances intact. Non-speech events longer than 60 seconds are excluded from the boundary computation, as they typically reflect persistent ambient conditions rather than discrete acoustic events; their annotations are still retained for downstream use. The remaining events are overlap-merged, expanded with short boundary padding, and cut at event gaps. A maximum segment length cap and a hard-cut fallback for very long recordings ensure training compatibility. The resulting segments are routed into the branch-specific annotation pipelines described below.

After segmentation, each segment retains its detected event labels. We map these fine-grained AudioSet labels into nine coarse-grained categories based on the AudioSet ontology: \textit{speech}, \textit{human voice (non-speech)}, \textit{singing}, \textit{music}, \textit{natural sounds}, \textit{source-ambiguous sounds}, \textit{sounds of things}, \textit{channel/environment/background}, and \textit{animal}. For each category, we compute the total duration within the segment using interval merging to avoid double-counting overlapping events. These per-segment category profiles determine which annotation branch each segment is routed to in the subsequent pipeline stages.

\subsection{ASR and Timestamp Alignment}
The ASR pipeline processes segments from the previous stage, retaining clips where "speech" or "singing" are among the detected tags. For each segment, we generate pseudo-labels using an ensemble of ASR systems, including models such as Qwen3-Omni-Instruct \cite{qwen3omni}, FunASR Nano \cite{an2025funasrtechnicalreport}, and Qwen3-ASR \cite{shi2026qwen3asr}. To ensure quality, the pipeline compares hypotheses across these systems and uses the inter-system word error rate (WER) as a consistency signal; segments with low cross-model WER are preserved as high-confidence data, while those with significant disagreement are discarded to minimize transcription noise. Furthermore, language identification (LID) is cross-validated using fastText \cite{joulin2017bag, joulin2016fasttext} on the recognized text and MMS-LID \cite{pratap2023mms} on the raw waveform, making the final language annotations robust against ASR errors, short utterances, and mixed-language cases.

For temporal grounding, we employ the TorchAudio MMS\_FA forced-alignment model \cite{JMLR:v25:23-1318} to synchronize the consensus-selected transcriptions with the audio waveform. This procedure generates precise word-level timestamps, which are subsequently aggregated into sentence-level segments during post-processing. This aggregation relies on punctuation detection and temporal boundary heuristics to ensure natural sentence breaks. Detailed examples of the resulting word-level and sentence-level serialization formats are provided in Appendix~\ref{sec:timestamp_examples}.

\subsection{Speech Caption}

The speech-caption branch depends on the event-preserving segmentation stage described in Section~\ref{sec:event_segmentation}. After segmentation, we use the sound event detection predictions to select segments that contain human vocal activity. A segment is routed to this branch when either its predicted speech score or singing score is greater than or equal to 0.5. For each selected segment, we apply DiariZen~\cite{han2025leveraging} to obtain speaker-aware segmentation, where each diarized region is associated with a speaker ID and a time interval. These diarized speaker regions serve as the basic units for speech-caption annotation, since voice attributes such as gender, age, accent, pitch, volume, speed, emotion, tone, and speaking style are speaker-dependent.

The speech-caption annotator is built in two stages. We first start from an internally trained single-speaker speech captioning model and apply it to the speaker-specific regions produced by DiariZen~\cite{han2025leveraging}, obtaining an initial collection of multi-speaker speech-caption data with speaker IDs, time spans, and speaker-level voice descriptions. Based on this bootstrapped data, we obtain the final multi-speaker speech captioning model, which is used as the speech-caption annotator in the MOSS-Audio pipeline. Given a vocal segment, this model produces speaker-aware captions that describe the acoustic and paralinguistic characteristics of different speakers, and the resulting annotations are later merged with other branch outputs into a unified text supervision target.

\subsection{Audio Caption}
The \textbf{general-audio branch} builds dense audio-caption for environmental sounds, open-domain acoustic scenes, and mixed real-world audio. It focuses on non-speech and mixed acoustic content, describing scene semantics, sound sources, vocal activity, event timelines, acoustic attributes, and temporal relations. The generated captions further serve as the semantic foundation for audio QA construction.

For real audio, we combine local event evidence with global semantic cues. PretrainedSED~\cite{schmid2024effectivepretrainingaudiotransformers} and Detect Any Sound~\cite{cai2025detectanysound} provide frame-level sound-event predictions under the AudioSet ontology, including event labels, sound sources, timestamps, and boundary information. These predictions are post-processed with event-type rules and temporal thresholds to obtain more reliable event metadata. In parallel, Qwen3-Omni-Captioner~\cite{qwen3omni} extracts global semantic anchors such as the overall scene, background atmosphere, and high-level audio summary.

Qwen3-Omni-30B-Thinking is then used as a fusion-based dense-caption generator. It integrates the global semantic anchors, post-processed event metadata, and the original audio to produce natural-language dense captions with acoustic attributes, foreground-background relations, source interactions, and temporal context. To improve reliability, candidate captions are verified against ASR annotations for speech regions, TimeAudio~\cite{wang2025timeaudio} outputs, and event metadata. An LLM-based judge further checks scene consistency, vocal activity, event correctness, source entities, acoustic attributes, and temporal coherence, and decides whether each sample should be kept, revised, or filtered.

In addition to real-audio annotation, we construct synthetic audio-caption data following Timestamped Audio Captioning (TAC)~\cite{kumar2026tactimestampedaudiocaptioning}. This path targets cases that are hard to annotate from real audio, such as rare event combinations, overlapping sounds, long-context transitions, and precise temporal boundaries. By composing audio from sound-effect libraries, environmental scenes, sound sources, and background audio with explicit event layouts and timestamps, the synthetic pipeline naturally provides controllable multi-granularity timestamped captions for dense-caption supervision.

\subsection{Music Caption}

The \textbf{music} branch is designed to convert raw musical audio into musically grounded supervision rather than generic audio descriptions. We first obtain a holistic base caption from an audio-language model, such as Qwen3-Omni~\cite{qwen3omni}, MusicFlamingo~\cite{ghosh2025musicflamingoscalingmusic}, or Audio-Flamingo~\cite{kong2024audioflamingo,ghosh2025audioflamingo2}. This caption provides high-level perceptual cues, including genre, production style, vocal presence, overall mood, and the global emotional trajectory of the track.

In parallel, the pipeline extracts symbolic and structural evidence with dedicated music-analysis tools. A MIR pipeline based on Chordino~\cite{mauch2010difficultchords}, BeatNet~\cite{heydari2021beatnet}, madmom~\cite{bock2016madmom}, Essentia~\cite{bogdanov2013essentia}, JukeMIR~\cite{castellon2021calm}, and related tools estimates chord sequences, beat and tempo statistics, key information, melody-related descriptors, and other low-level musical attributes. An instrument-recognition branch records time-varying active instruments, while SongFormer~\cite{hao2026songformerscalingmusicstructure} predicts the song structure and divides each track into musically meaningful regions such as intro, verse, chorus, bridge, instrumental, and outro. These structural boundaries are then used to cut the original track into segment-level clips. For each segment, we run lyrics ASR when vocals are present and perform segment-level key analysis, yielding aligned tuples of \emph{structure label, timestamp, key, chord progression, and lyrics}.


The final music caption is generated by an instruction LLM from the merged music metadata. Its prompt is explicitly constrained to synthesize a coherent listener-facing description covering style or genre, tempo feel, tonal center, harmonic movement, instrumentation, production texture, vocal and lyrical content when available, structural development, dynamics, and mood. To avoid tool artifacts in the target text, the generator is forbidden from mentioning field names, JSON keys, Lyrics, metadata, or intermediate analysis tools. It is also instructed to trust the holistic ALM caption when weak or missing segment-level evidence could otherwise lead to false claims, for example treating missing lyric transcripts as proof that a track is instrumental. This produces natural captions that preserve specialist music information while remaining suitable as unified autoregressive training targets.



\subsection{Caption Merge \& Refine}
A single audio clip typically carries multiple annotation branches—ASR transcripts, speaker-attribute descriptions, dense audio captions, scene-level summaries, and  music or acoustic analyses—each produced by a different upstream specialist.
Since holistic audio captioning is a primary pretraining task, these branch-specific annotations must be consolidated into a unified caption target.

\textbf{Canonical normalization.}
All upstream annotations are first projected into a unified \texttt{tool\_results} interface, where heterogeneous outputs are organized into logical slots such as \texttt{asr}, \texttt{event\_caption},  \texttt{speech\_caption}, and \texttt{music\_caption}.
This abstraction decouples the downstream merge procedure from dataset-specific schemas and allows evidence from different annotation pipelines to be consumed in a consistent format.
In parallel, global audio-class scores are normalized to the range $[0,1]$ and further aggregated into three coarse prior axes: \emph{speech}, \emph{music}, and \emph{event}.
These priors provide a compact estimate of the dominant acoustic content of each clip and serve as routing signals for subsequent evidence selection.

\textbf{Prior-driven routing.}
Given the normalized priors and top-$k$ class predictions, a lightweight routing policy, Router-R1, determines which evidence branches should be included in the merged target and specifies their relative ordering.
The routing policy estimates modality dominance while adopting deliberately conservative thresholds, so that weak but semantically meaningful signals, such as low-energy speech or faint background music, are preserved whenever possible.
To account for ambiguous clips, residual uncertainty is measured using the entropy of the class distribution and incorporated into the routing decision.
The router also applies a set of quality-control constraints: empty or highly repetitive ASR hypotheses are removed; speech-related evidence is excluded when the speech prior is negligible; non-linguistic human vocalizations are prevented from being treated as lexical speech; and music-related claims from the general captioner are suppressed when the specialized music branch indicates that music is absent.
These constraints reduce hallucinated or redundant evidence before final synthesis.

\textbf{Constrained synthesis.}
The selected evidence is converted into the final caption target through a two-stage LLM-based synthesis protocol.
First, a planning prompt produces a structured JSON object containing the primary theme, selected evidence sources, merge order, and rationale, with the constraint that only non-empty slots may be referenced.
Second, a generation prompt synthesizes a single English description from the planned evidence.
The generated target is required to preserve available timestamps, speaker attributes, and event chronology, while omitting information from absent or filtered branches.
In addition to the LLM-generated target, a deterministic fallback target following the same evidence ordering is produced to improve robustness against unstable generation.
The final output is therefore a unified and information-dense caption target for holistic audio captioning.
\section{Pretraining}

The goal of pretraining is to establish a robust audio--language alignment before the model is exposed to complex instruction-following and reasoning tasks. Without a well-aligned audio prefix, later stages of supervised fine-tuning and reinforcement learning cannot effectively teach the model to interpret acoustic content. We therefore organize the pretraining data into three objective groups that jointly build this alignment: \textbf{ASR-related tasks} for precise audio-to-text transcription, \textbf{audio captioning} for open-ended audio understanding, and \textbf{text-only language modeling} to preserve the decoder's general language capability. The default sampling ratio is 30\% for ASR-related tasks, 40\% for audio captioning, and 30\% for text-only language modeling. Overall, the pretraining stage uses approximately 1.2T training tokens.


The \textbf{ASR-related tasks} include ordinary ASR, word-level timestamp ASR, and sentence-level timestamp ASR. Ordinary ASR trains the model to transcribe spoken content from audio. Word-level timestamp ASR adds fine-grained temporal supervision by associating recognized words with their timestamps. Sentence-level timestamp ASR uses sentence segments with start and end times, providing a more stable form of temporal alignment. These tasks are mixed together as the ASR-related pretraining pool.

The \textbf{audio captioning objective} uses the final merged captions produced by our caption construction pipeline. For each segmented clip, the pipeline first collects available evidence from different annotation branches, such as ASR, speech-related descriptions, music-related descriptions, and general audio descriptions. These branch outputs are then merged into a unified natural-language caption. The model is trained to generate this merged caption for the basic capability of  understanding all kinds of audio in the real world.

The \textbf{text-only language modeling objective} uses high-quality text pretraining data without instruction-style formatting. This corpus covers a broad range of domains, including mathematics, code, education, literature, and general text. It is included to preserve the decoder's original text modeling capability during audio-language pretraining. Since the model is exposed to large amounts of audio-conditioned data, mixing pure text pretraining data helps prevent degradation of general language ability, such as fluent generation, knowledge expression, reasoning over text, and code understanding. By default, this objective is enabled in the fully opened training stage, where the language model is jointly updated together with the audio encoder, adapter, and DeepStack modules.

Within each objective group, different datasets are sampled with a square-root mixing strategy. Instead of sampling datasets strictly according to their raw sizes, we assign each dataset a probability proportional to the square root of its size. This reduces the dominance of very large datasets while still allowing larger datasets to contribute more samples than smaller ones.

Pretraining is conducted in two stages.
In \textbf{Stage 1}, training mainly focuses on the modality adapter and the DeepStack cross-layer injection modules, while the audio encoder and language model are kept relatively stable. Since this stage is used to open and stabilize the audio-prefix pathway, text-only data is not mixed by default. The training mixture therefore contains only audio-text objectives, including ASR-related tasks and audio captioning.
In \textbf{Stage 2}, the full model is optimized end to end under the complete objective mixture. The audio encoder, modality adapter, DeepStack injection modules, and language model are jointly updated. Text-only data is enabled in this stage, so the model is trained with the full mixture of ASR-related tasks, audio captioning, and text-only language modeling.



\section{Post-Training}

After pretraining establishes the basic audio--language interface, MOSS-Audio undergoes staged post-training to produce two distinct model variants: instruction-following models that execute user requests directly and accurately, and reasoning-capable models that perform structured multi-step analysis over audio content. The post-training process consists of three phases: supervised fine-tuning for task adaptation, reasoning cold start for thinking-pattern initialization, and reinforcement learning for robustness improvement.

\subsection{Supervised Fine-Tuning}

The first post-training stage is supervised fine-tuning, which adapts the pretrained model to user-facing instruction formats and diverse audio-centered tasks. The SFT mixture consists of audio question answering data, captioning data, ASR and timestamp ASR data and self-identity data. The QA data is generated by language models from speech captions, music captions, and general audio captions, covering tasks such as speech understanding, speaker attribute analysis, acoustic event understanding, music understanding, temporal reasoning, and scene-level audio comprehension. The captioning data includes overall captions, speech captions, music captions, and general audio captions. The ASR data includes ordinary transcription, word-level timestamp ASR, and sentence-level timestamp ASR. Self-identity data is added to standardize responses about the model's name, developer, capabilities, and limitations.

This stage trains MOSS-Audio to follow natural instructions, produce task-specific output formats, and respond consistently across different kinds of questions. It produces the instruction-following variants of MOSS-Audio.

\subsection{Reasoning Cold Start}

To obtain the reasoning-oriented variants, we introduce a reasoning cold-start stage after supervised fine-tuning. This stage initializes the model with stable reasoning behavior before reinforcement learning. The cold-start mixture includes both audio-centered reasoning data and text-only reasoning data.

The audio-centered reasoning data teaches the model to connect final answers with audio-relevant evidence, such as spoken content, speaker attributes, prosody, emotion, acoustic events, temporal relations, music structure, instrumentation, vocal characteristics, and lyrics. These samples encourage the model to organize perceptual evidence and perform multi-step analysis for complex audio understanding tasks.

The text-only reasoning data is added to transfer general reasoning patterns from the text modality to audio-language modeling. Although these samples do not contain audio inputs, they help the model acquire a more stable reasoning paradigm that can later be applied to audio-centered tasks.

This stage improves the model's ability to perform structured reasoning before reinforcement learning. Compared with standard SFT, reasoning cold start focuses less on task-format adaptation and more on evidence-grounded analysis, multi-step reasoning, and reliable final-answer generation.

\subsection{Reinforcement Learning}

After the model has acquired basic reasoning behavior through cold-start supervision, we further optimize it with reinforcement learning. We use a DAPO-based reinforcement learning stage to improve answer correctness, reasoning robustness, and format compliance across diverse audio tasks.

The RL data covers multiple audio domains, including speech understanding, paralinguistic analysis, environmental sound understanding, music understanding, audio question answering, and temporal reasoning. For each prompt, the model samples multiple responses. These responses are evaluated using rewards that reflect task correctness, response quality, format compliance, and the usefulness of the reasoning process. The policy is then updated to increase the probability of higher-reward responses.

Compared with the cold-start stage, which imitates teacher-provided reasoning traces, the reinforcement learning stage optimizes the model through online sampling and reward-based comparison. This helps the model move beyond fixed reasoning templates and improves its robustness on more diverse and difficult audio understanding problems.

For rollout generation, we sample responses with a temperature of $1.0$, top-$p$ of $1.0$ , and top-$k$ of $50$ . Each prompt is expanded into $16$ sampled responses, and the rollout batch size is set to $128$. The maximum response length is limited to $2048$ tokens, which provides sufficient space for reasoning while preventing excessively long generations. In each training round, we use $160$ rollouts for policy optimization.

For policy optimization, we adopt a clipped DAPO objective. The lower clipping coefficient is set to $\epsilon=0.2$ , while the higher clipping coefficient is set to $\epsilon_{\mathrm{high}}=0.28$ . This asymmetric clipping allows slightly more aggressive updates for beneficial trajectories while still constraining unstable policy shifts. We additionally enable token-level importance sampling correction, with the TIS clipping threshold set to $2.0$ and the lower clipping bound set to $0.0$ . This correction stabilizes training when rollout trajectories are generated by a policy that may differ from the current policy being updated.


During DAPO training, we apply dynamic filtering to discard rollout groups whose reward standard deviation is (near) zero. Such groups arise when all sampled responses of a prompt receive the same reward---typically all-correct or all-wrong rollouts---and therefore yield no within-group advantage signal under the group-relative objective. Filtering them out keeps each update focused on prompts where positive and negative trajectories coexist, improving the efficiency of policy optimization.

\begin{figure}[htbp]
    \centering
    \includegraphics[width=0.65\linewidth]{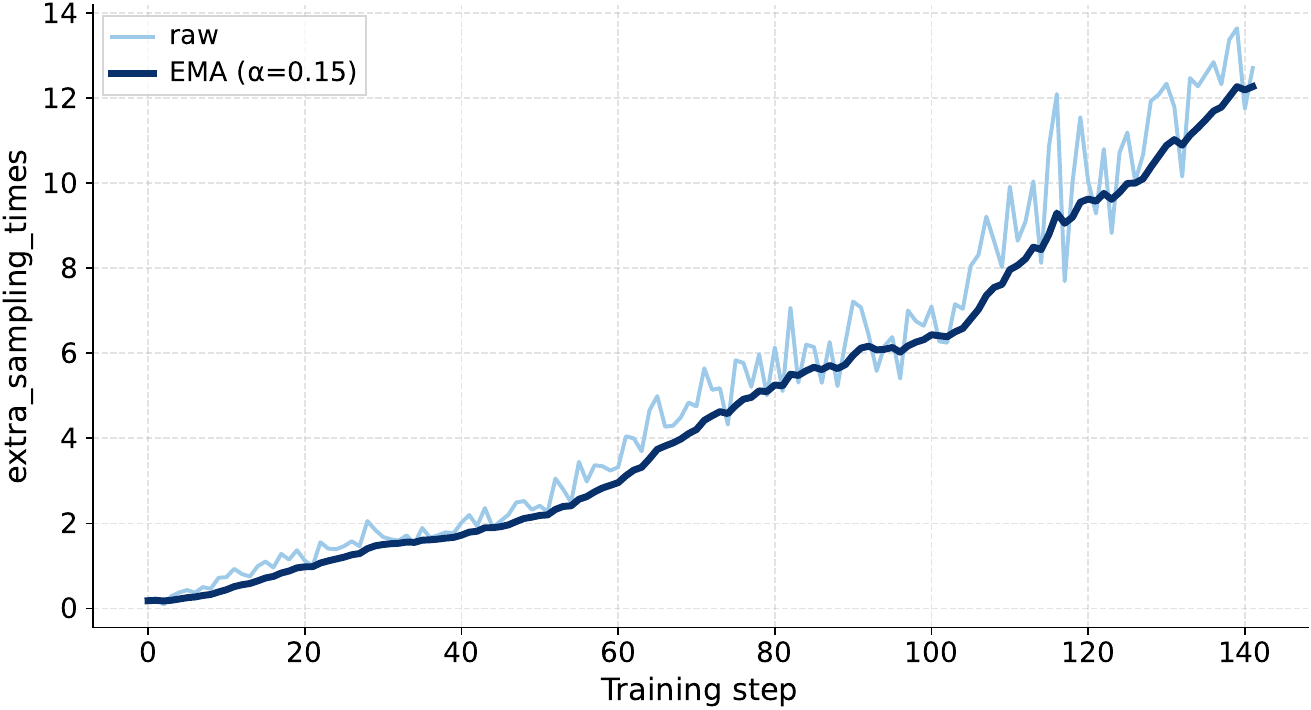}
    \caption{ Extra over-sampling triggered by dynamic filtering during DAPO training. At each step, rollout groups with zero reward standard deviation are discarded because they carry no within-group advantage signal, and extra over-sampling rounds are issued to refill the batch. The curve reports the number of these extra over-sampling rounds per step; the light curve is the step-level value and the dark curve its exponential moving average ($\alpha=0.15$).}
    \label{fig:rl-drop-zero-std}
\end{figure}


As shown in Figure~\ref{fig:rl-drop-zero-std}, the number of extra over-sampling rounds triggered by dynamic filtering grows steadily throughout training, rising from $1$ to a maximum of $14$ at step $139$. This means an increasing share of sampled groups is filtered out as zero-std---predominantly because the model comes to solve these prompts on every rollout---so progressively more over-sampling is needed to assemble a full batch of informative groups. Equivalently, the effective learning signal concentrates on the shrinking set of prompts with non-trivial reward variance. This is precisely why dynamic filtering matters: it keeps DAPO updates focused on prompts where positive and negative trajectories coexist, at the cost of the additional rollout sampling that this curve quantifies.



\begin{figure}[htbp]
      \centering

      \begin{subfigure}{0.48\linewidth}
          \centering
          \includegraphics[width=\linewidth]{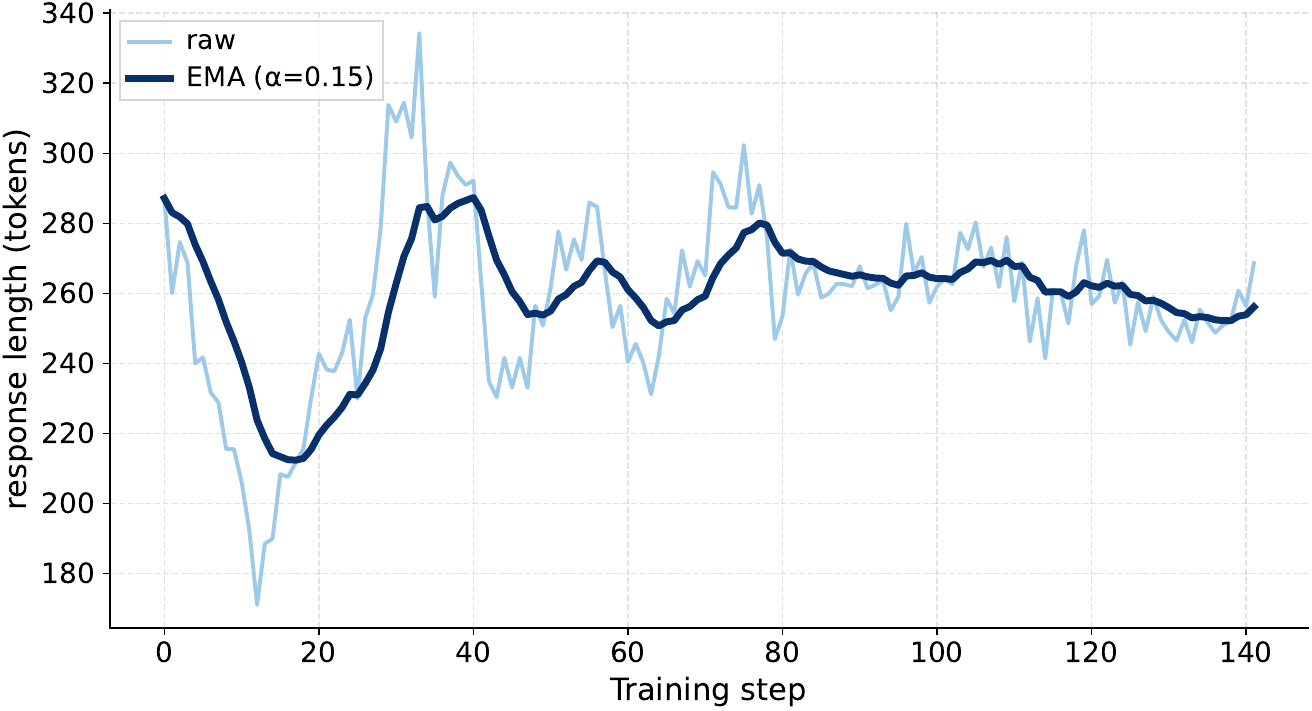}
          \caption{Response length}
          \label{fig:rl-response-length}
      \end{subfigure}
      \hfill
      \begin{subfigure}{0.48\linewidth}
          \centering
          \includegraphics[width=\linewidth]{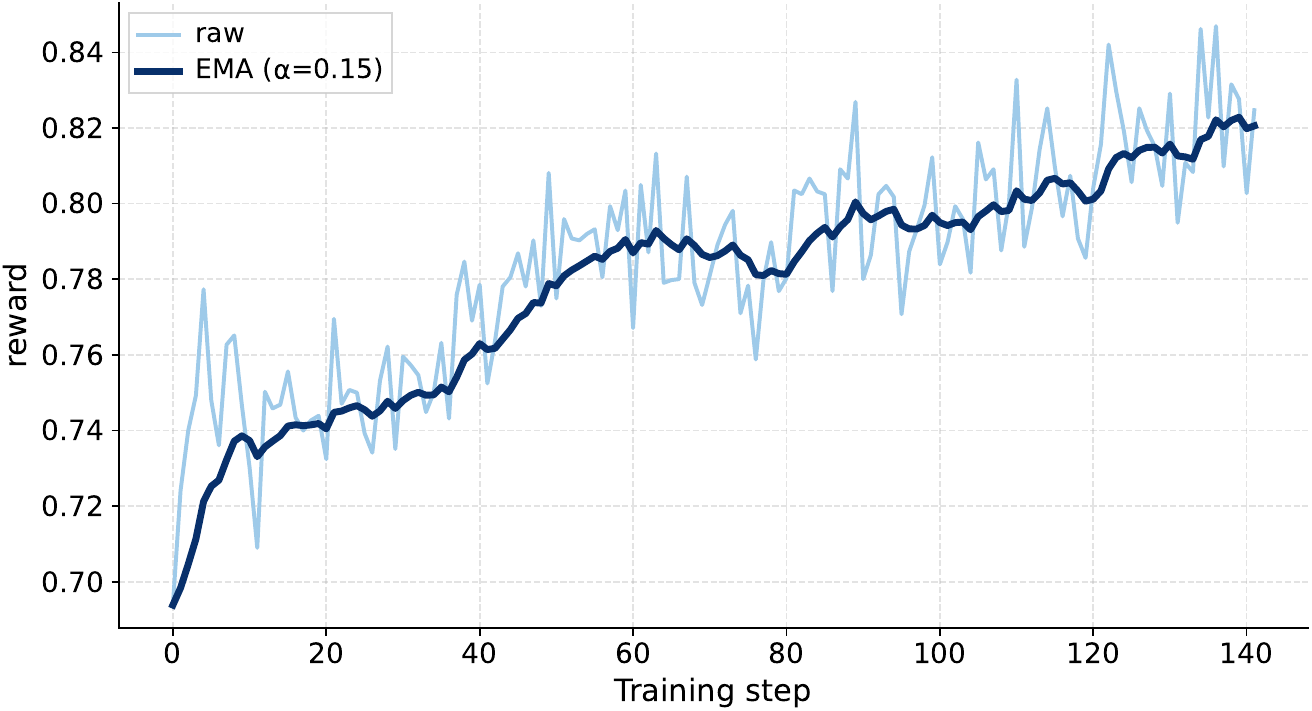}
          \caption{Rollout raw reward}
          \label{fig:rl-raw-reward}
      \end{subfigure}

      \caption{
      Evolution of response length and rollout raw reward during DAPO training.
      The response length becomes stable after early-stage fluctuations, while the rollout reward increases steadily, suggesting that the model improves reward without relying on excessively long generations.
      }
      \label{fig:rl-training-dynamics}
  \end{figure}

Figure~\ref{fig:rl-response-length} and Figure~\ref{fig:rl-raw-reward} presents the evolution of rollout reward and response length during DAPO training. The left plot shows a steady improvement in rollout raw reward. The EMA curve rises from approximately $0.69$ at the beginning to above $0.82$ near the end of training, with the maximum raw reward reaching $0.847$ at step $136$. This trend indicates that DAPO effectively improves the policy, enabling the model to generate responses that better satisfy the reward criteria, including answer correctness, response quality, and format compliance.

The right plot shows the corresponding response length dynamics. Unlike the reward curve, the response length does not increase monotonically. It first drops sharply in the early stage, reaching a minimum of $171.065$ tokens at step $12$, then temporarily increases and peaks at $334.324$ tokens around step $33$. After this transient fluctuation, the EMA curve gradually stabilizes around $250$--$270$ tokens. This suggests that the model does not obtain higher rewards simply by producing longer responses. Instead, after the initial exploration stage, DAPO encourages more effective and compact reasoning behavior.

Combining the two curves, we observe a desirable training pattern: the reward continues to improve while the average response length remains controlled. This indicates that the model learns to improve answer quality and reasoning reliability without relying on unnecessarily redundant long outputs. Such behavior is particularly important for audio reasoning tasks, where overly long reasoning may introduce hallucinated acoustic evidence or reduce response efficiency. Therefore, the reward and length curves jointly demonstrate that the DAPO stage improves both task performance and generation stability.

Compared with purely supervised distillation, the DAPO stage encourages the model to explore better reasoning trajectories and improves its robustness across different audio domains. It also helps reduce common failure modes, such as hallucinated acoustic evidence, text-only surrogate reasoning, unstable output formatting, and unnecessarily redundant thinking. By combining thinking-process distillation with domain-diverse DAPO optimization, the post-training pipeline enables the model to produce reasoning that is not only structurally coherent, but also better aligned with the actual audio input.

\section{Evaluation}
\label{sec:evaluation}

We evaluate MOSS-Audio on four groups of tasks: \textbf{general audio understanding}, \textbf{speech captioning}, \textbf{automatic speech recognition (ASR)}, and \textbf{timestamp-aware ASR}. These evaluations cover both high-level audio comprehension and speech-centric perception. General audio understanding measures the model's ability to answer questions about speech, music, sound events, and acoustic scenes. Speech captioning evaluates fine-grained description of speaker and utterance attributes. ASR measures transcription accuracy under diverse speech conditions, while timestamp ASR further evaluates whether the model can align recognized content with time.

\subsection{General Audio Understanding}
\label{subsec:general_audio_understanding}

We evaluate general audio understanding on \textbf{MMAU}, \textbf{MMAU-Pro}, \textbf{MMAR}, and \textbf{MMSU}. We report the arithmetic average over the four benchmarks as the main aggregate score, and compare MOSS-Audio with representative open-source and proprietary audio-language models.

\begin{figure}[t]
  \centering
  \begin{subfigure}[t]{0.53\linewidth}
    \centering
    \includegraphics[width=\linewidth]{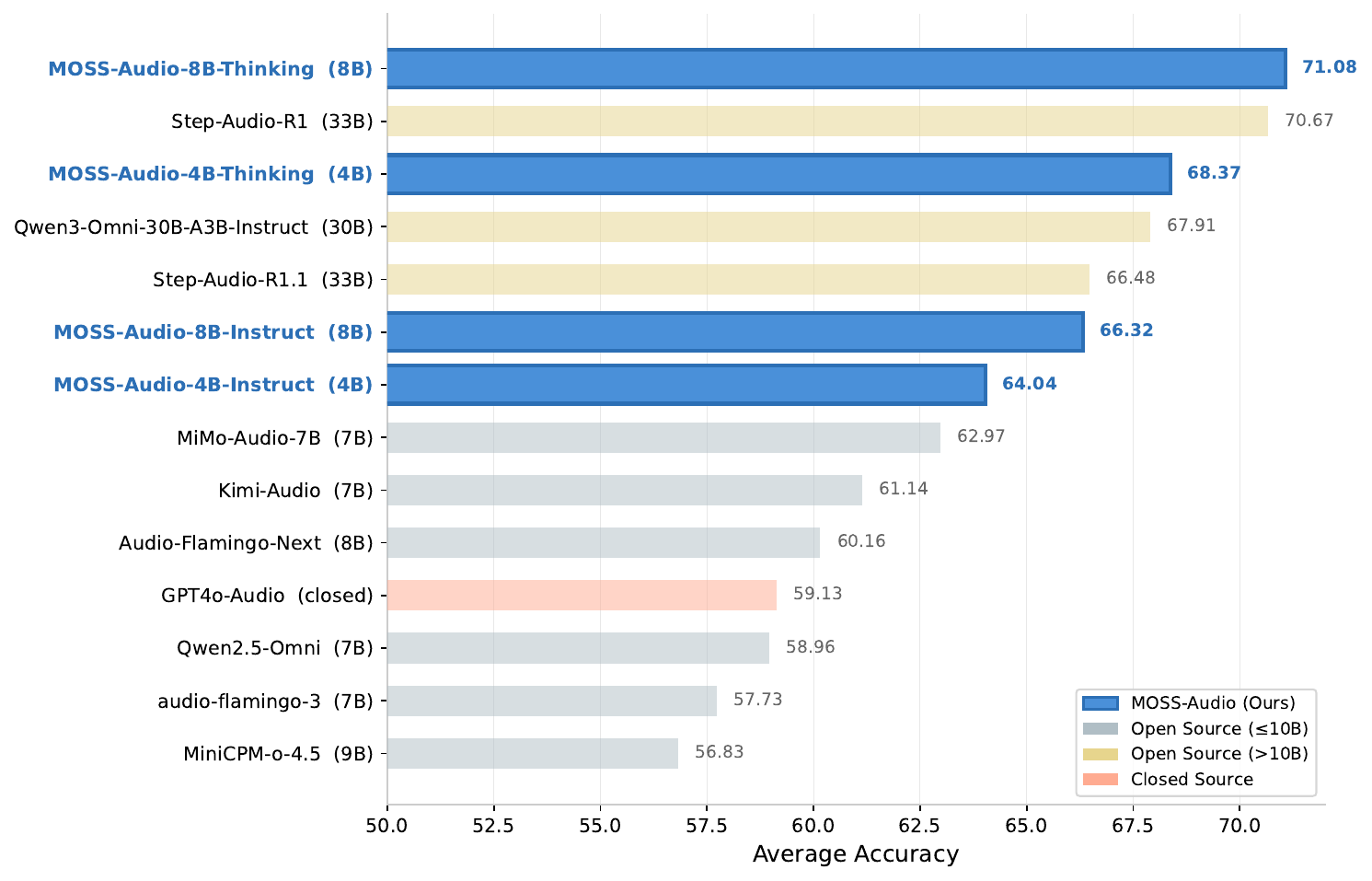}
    \caption{General audio understanding results on MMAU, MMAU-Pro, MMAR, and MMSU.}
    \label{fig:general_audio_understanding}
  \end{subfigure}
  \hfill
  \begin{subfigure}[t]{0.45\linewidth}
    \centering
    \includegraphics[width=\linewidth]{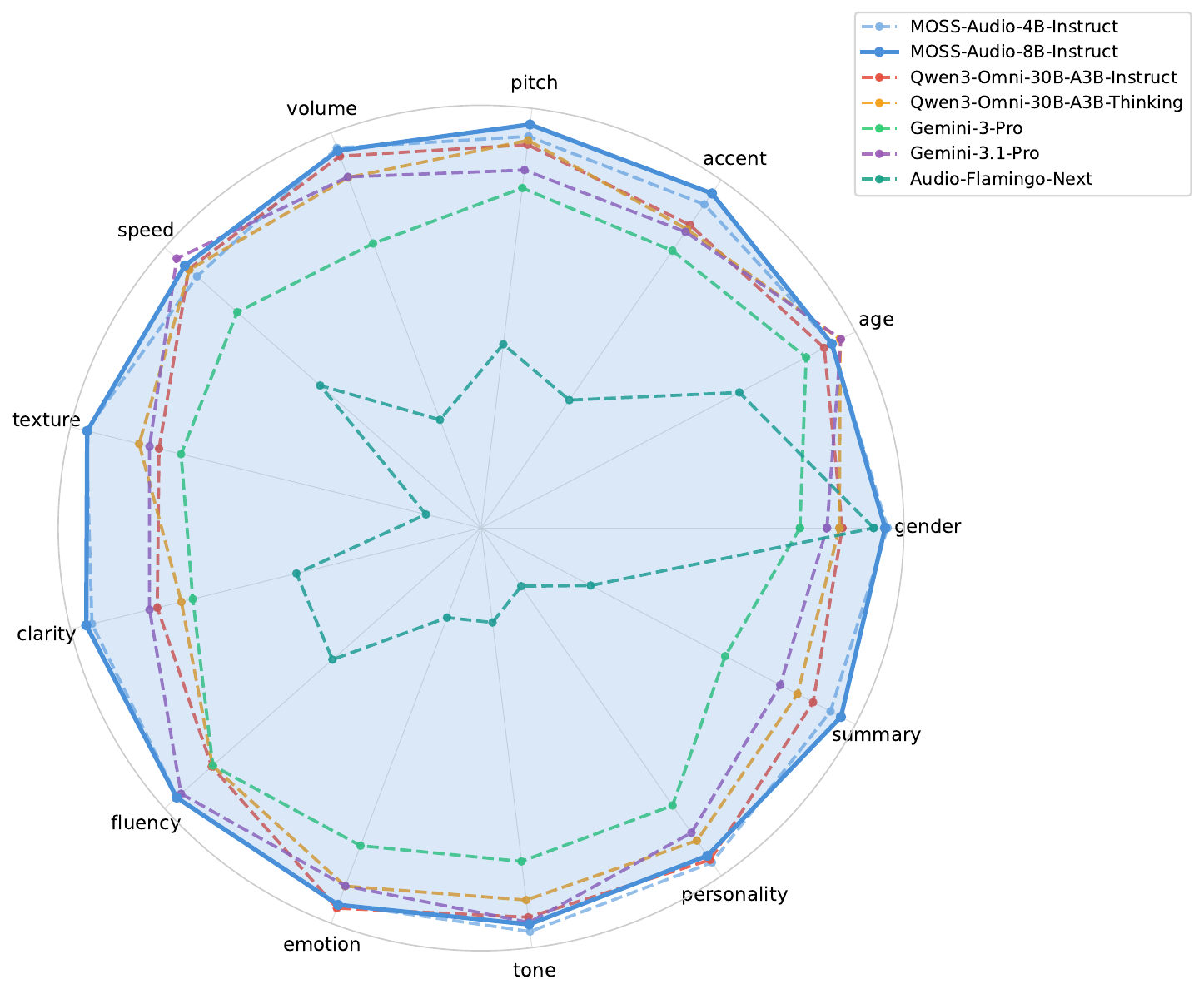}
    \caption{Speech captioning results across 13 judged dimensions.}
    \label{fig:speech_captioning}
  \end{subfigure}
  \caption{Evaluation visualizations. The left plot summarizes performance across four general audio understanding benchmarks, while the right plot shows fine-grained speech captioning behavior across 13 judged dimensions.}
  \label{fig:evaluation_visualization}
\end{figure}

\begin{table*}[htbp]
\centering
\caption{\textbf{General audio understanding results on MMAU, MMAU-Pro, MMAR, and MMSU.} Higher scores are better.}
\label{tab:general_audio_understanding}
\small
\setlength{\tabcolsep}{5pt}
\renewcommand{\arraystretch}{1.08}
\begin{tabular}{@{}lcccccc@{}}
\toprule
\multicolumn{1}{c}{\textbf{Model}} & \textbf{Size} & \textbf{MMAU} & \textbf{MMAU-Pro} & \textbf{MMAR} & \textbf{MMSU} & \textbf{Avg} \\
\midrule
\multicolumn{7}{c}{\textit{Closed-source models}} \\
\midrule
GPT4o-Audio & -- & 65.66 & 52.30 & 59.78 & 58.76 & 59.13 \\
Gemini-3-Pro & -- & 80.15 & 68.28 & 81.73 & 81.28 & 77.86 \\
Gemini-3.1-Pro & -- & \textbf{81.10} & \textbf{73.47} & \textbf{83.70} & \textbf{81.30} & \textbf{79.89} \\
\midrule
\multicolumn{7}{c}{\textit{Open-source models}} \\
\midrule
Qwen3-Omni-30B-A3B-Instruct & 30B & 75.00 & 61.22 & 66.40 & 69.00 & 67.91 \\
Step-Audio-R1.1 & 33B & 72.18 & 60.80 & 68.75 & 64.18 & 66.48 \\
Step-Audio-R1 & 33B & \textbf{78.67} & 59.68 & \textbf{69.15} & 75.18 & 70.67 \\
Kimi-Audio & 7B & 72.41 & 56.58 & 60.82 & 54.74 & 61.14 \\
Qwen2.5-Omni & 7B & 65.60 & 52.20 & 56.70 & 61.32 & 58.96 \\
Audio Flamingo 3 & 7B & 61.23 & 51.70 & 57.96 & 60.04 & 57.73 \\
Audio Flamingo Next & 8B & 76.10 & 56.34 & 51.01 & 57.20 & 60.16 \\
MiMo-Audio-7B & 7B & 74.90 & 53.35 & 61.70 & 61.94 & 62.97 \\
MiniCPM-o-4.5 & 9B & 70.97 & 39.65 & 55.75 & 60.96 & 56.83 \\
\textbf{MOSS-Audio-4B-Instruct} & 4B & 75.79 & 58.16 & 62.54 & 59.68 & 64.04 \\
\textbf{MOSS-Audio-4B-Thinking} & 4B & 77.64 & 60.75 & 63.91 & 71.20 & 68.37 \\
\textbf{MOSS-Audio-8B-Instruct} & 8B & 77.03 & 57.48 & 64.42 & 66.36 & 66.32 \\
\textbf{MOSS-Audio-8B-Thinking} & 8B & 77.33 & \textbf{64.92} & 66.53 & \textbf{75.52} & \textbf{71.08} \\
\bottomrule
\end{tabular}
\end{table*}

As shown in Table~\ref{tab:general_audio_understanding} and Figure~\ref{fig:general_audio_understanding}, MOSS-Audio achieves the best performance among all open-source models in this evaluation group. \textbf{MOSS-Audio-8B-Thinking} obtains the highest open-source average score of \textbf{71.08} across MMAU, MMAU-Pro, MMAR, and MMSU, establishing the strongest overall result among the compared open-source audio-language models.

A notable pattern is that MOSS-Audio achieves this result with a compact model size. \textbf{MOSS-Audio-4B-Thinking} already outperforms several larger 8B-scale open-source baselines, while \textbf{MOSS-Audio-8B-Thinking} further surpasses a number of substantially larger models, including 30B-scale models. 

The Thinking variants consistently outperform their paired Instruct variants at both 4B and 8B scales, showing that the reasoning-oriented branch is more suitable for broad audio understanding tasks. While proprietary models still remain strong references, MOSS-Audio sets the leading open-source result under this benchmark suite and demonstrates favorable scaling efficiency across both released model sizes.

\subsection{Speech Captioning}
\label{subsec:speech_captioning}

Speech captioning evaluates whether a model can generate faithful natural-language descriptions of speech content and paralinguistic information. To support this evaluation, we construct a dedicated speech captioning benchmark with 2,000 speech audio samples. We first use audio-language models to annotate each candidate audio with preliminary speaker-related tags, covering attributes such as gender, age, accent, pitch, volume, speaking speed, voice texture, clarity, fluency, emotion, tone, personality, and utterance summary. Based on these tags, we then perform balanced sampling across the major categories of each dimension, so that the final benchmark covers diverse speakers, acoustic conditions, speaking styles, and affective states. This produces a domain-balanced evaluation set that is suitable for measuring fine-grained speech captioning ability.

For each selected audio sample, human annotators write reference captions along 13 judged dimensions. The annotations are further reviewed through a strict quality-control process to ensure that the references are accurate, dimension-specific, and grounded in the audio.

During evaluation, each model is prompted to generate speech captions in the same 13-dimensional format. For each dimension, we provide both the model output and the human reference to a text-based judge model, which scores how well the model prediction matches the reference description. The final score of each model is computed by averaging the dimension-level matching scores over all evaluated samples.

\providecommand{\mossaudiomodelheader}[1]{\shortstack[c]{#1}}
\begin{table*}[t]
\centering
\caption{\textbf{Speech captioning results across 13 judged dimensions.} Rows correspond to judged dimensions and columns correspond to models. Higher is better.}
\label{tab:speech_captioning}
\small
\setlength{\tabcolsep}{3pt}
\renewcommand{\arraystretch}{1.10}
\resizebox{\textwidth}{!}{%
\begin{tabular}{@{}lccccccc@{}}
\toprule
\textbf{Dimension} & \mossaudiomodelheader{\textbf{Audio-Flamingo}\\\textbf{-Next}} & \mossaudiomodelheader{\textbf{Qwen3-Omni}\\\textbf{-Instruct}} & \mossaudiomodelheader{\textbf{Qwen3-Omni}\\\textbf{-Thinking}} & \mossaudiomodelheader{\textbf{Gemini-3}\\\textbf{-Pro}} & \mossaudiomodelheader{\textbf{Gemini-3.1}\\\textbf{-Pro}} & \mossaudiomodelheader{\textbf{MOSS-Audio}\\\textbf{-4B-Instruct}} & \mossaudiomodelheader{\textbf{MOSS-Audio}\\\textbf{-8B-Instruct}} \\
\midrule
Gender & 4.617 & 4.436 & 4.419 & 4.191 & 4.347 & \textbf{4.697} & 4.683 \\
Age & 3.461 & 3.936 & 4.026 & 3.835 & \textbf{4.030} & 3.980 & 3.979 \\
Accent & 3.160 & 4.356 & 4.327 & 4.181 & 4.310 & 4.497 & \textbf{4.572} \\
Pitch & 2.679 & 3.590 & 3.610 & 3.392 & 3.474 & 3.628 & \textbf{3.682} \\
Volume & 2.391 & 3.682 & 3.577 & 3.254 & 3.580 & \textbf{3.722} & 3.709 \\
Speed & 2.818 & 3.614 & 3.610 & 3.320 & \textbf{3.687} & 3.564 & 3.638 \\
Texture & 1.941 & 3.093 & 3.179 & 2.998 & 3.134 & \textbf{3.407} & 3.403 \\
Clarity & 2.839 & 3.521 & 3.403 & 3.347 & 3.559 & 3.841 & \textbf{3.869} \\
Fluency & 2.788 & 3.531 & 3.526 & 3.524 & 3.720 & 3.744 & \textbf{3.747} \\
Emotion & 2.056 & \textbf{3.328} & 3.232 & 3.055 & 3.231 & 3.311 & 3.314 \\
Tone & 2.025 & 3.224 & 3.154 & 2.997 & 3.245 & \textbf{3.282} & 3.253 \\
Personality & 1.940 & 3.292 & 3.197 & 3.023 & 3.158 & \textbf{3.305} & 3.272 \\
Summary & 2.157 & 3.179 & 3.107 & 2.775 & 3.028 & 3.259 & \textbf{3.307} \\
\midrule
\textbf{Average} & 2.683 & 3.599 & 3.567 & 3.376 & 3.577 & 3.711 & \textbf{3.725} \\
\bottomrule
\end{tabular}%
}
\end{table*}

As shown in Table~\ref{tab:speech_captioning} and Figure~\ref{fig:speech_captioning}, MOSS-Audio achieves the best overall speech captioning performance among all compared models, including strong proprietary systems such as Gemini-3.1-Pro. \textbf{MOSS-Audio-8B-Instruct} obtains the highest average score of \textbf{3.7252}, followed closely by \textbf{MOSS-Audio-4B-Instruct} with \textbf{3.7105}. This shows that MOSS-Audio is highly effective at fine-grained speech description, especially for speaker attributes, prosodic cues, voice quality, speaking style, and utterance-level summarization.

\subsection{ASR}
\label{subsec:asr}

We evaluate ASR across 12 dimensions, including health-condition speech, dialectal speech, singing, non-speech vocalizations, code-switching, clean and noisy environments, whisper speech, far-field and near-field audio, multi-speaker audio, age-related subsets, and semantic-content subsets. We report character error rate (CER), where lower is better. Detailed dataset-level results are provided in Appendix~\ref{sec:complete_asr_results}.

\providecommand{\mossaudiomodelheader}[1]{\shortstack[c]{#1}}
\begin{table*}[htbp]
\centering
\caption{\textbf{ASR summary results across 12 evaluation dimensions.} Rows correspond to evaluation dimensions and columns correspond to models. Lower CER is better.}
\label{tab:asr_summary}
\scriptsize
\setlength{\tabcolsep}{0.9pt}
\renewcommand{\arraystretch}{1.10}
\resizebox{\textwidth}{!}{%
\begin{tabular}{@{}>{\raggedright\arraybackslash}p{2.55cm}ccccccccccc@{}}
\toprule
\multicolumn{1}{c}{\textbf{Dimension}} & \mossaudiomodelheader{\textbf{Paraformer}\\\textbf{-Large}} & \mossaudiomodelheader{\textbf{GLM-ASR}\\\textbf{-Nano}} & \mossaudiomodelheader{\textbf{Fun-ASR}\\\textbf{-Nano}} & \mossaudiomodelheader{\textbf{SenseVoice}\\\textbf{-Small}} & \mossaudiomodelheader{\textbf{Kimi-Audio}\\\textbf{-7B-Instruct}} & \mossaudiomodelheader{\textbf{Audio-Flamingo}\\\textbf{-Next}} & \mossaudiomodelheader{\textbf{Qwen2.5}\\\textbf{-Omni-3B}} & \mossaudiomodelheader{\textbf{Qwen2.5}\\\textbf{-Omni-7B}} & \mossaudiomodelheader{\textbf{Qwen3-Omni}\\\textbf{-Instruct}} & \mossaudiomodelheader{\textbf{MOSS-Audio}\\\textbf{-4B-Instruct}} & \mossaudiomodelheader{\textbf{MOSS-Audio}\\\textbf{-8B-Instruct}} \\
\midrule
Health Condition & 22.18 & 24.49 & 21.99 & 24.04 & 21.11 & 36.13 & 24.65 & 23.85 & 20.73 & 21.11 & \textbf{19.18} \\
Dialect & 43.45 & 22.39 & \textbf{7.80} & 8.89 & 29.34 & 25.85 & 33.87 & 31.91 & 15.63 & 11.84 & 8.76 \\
Singing & 32.34 & 51.95 & 19.35 & 23.79 & 21.76 & 25.80 & 24.24 & 22.69 & 16.01 & 10.79 & \textbf{9.81} \\
Non-Speech Vocal. & 4.95 & 4.65 & 4.76 & 4.92 & 4.68 & 8.25 & 5.54 & 4.56 & 4.73 & \textbf{4.01} & 4.31 \\
Code-Switching & 12.65 & 11.88 & 11.23 & 13.90 & 16.38 & 34.53 & 11.66 & 12.97 & 11.30 & \textbf{10.11} & 10.18 \\
Acoustic Env. (Clean) & 3.11 & 3.68 & 2.98 & 4.13 & \textbf{2.20} & 8.64 & 2.76 & 2.52 & 2.23 & 3.11 & 2.70 \\
Acoustic Env. (Noisy) & 4.67 & 5.02 & 3.46 & 4.93 & \textbf{2.15} & 38.85 & 3.56 & 3.16 & 2.47 & 3.72 & 3.20 \\
Whisper & 5.02 & 4.94 & 3.78 & 5.57 & 2.66 & 12.84 & 4.32 & 3.64 & \textbf{1.90} & 3.29 & 2.75 \\
Far-Field / Near-Field & 17.46 & 27.51 & 18.38 & 26.66 & 21.02 & 42.90 & 22.15 & 25.38 & \textbf{17.08} & 18.48 & 24.04 \\
Multi-Speaker & 20.33 & 28.02 & 19.82 & 24.06 & 20.61 & 62.13 & 22.91 & 21.01 & \textbf{18.15} & 20.33 & 24.36 \\
Age & 14.96 & 17.19 & 14.95 & 17.63 & 16.74 & 38.30 & 15.17 & 16.13 & \textbf{11.46} & 15.09 & 15.26 \\
Semantic Content & 7.14 & 7.32 & 6.08 & 7.55 & 6.12 & 30.18 & 7.24 & 6.78 & \textbf{5.74} & 8.15 & 7.69 \\
\midrule
\textbf{Average} & 15.77 & 17.29 & 12.04 & 14.50 & 14.12 & 30.19 & 15.26 & 15.05 & 11.39 & 11.58 & \textbf{11.30} \\

\bottomrule
\end{tabular}%
}
\end{table*}

As shown in Table~\ref{tab:asr_summary}, \textbf{MOSS-Audio-8B-Instruct} achieves the best overall CER of \textbf{11.30}, followed by Qwen3-Omni-30B-A3B-Instruct and \textbf{MOSS-Audio-4B-Instruct}. This shows that MOSS-Audio preserves strong transcription accuracy while retaining broader audio-language capabilities.

\subsection{Timestamp ASR}
\label{subsec:timestamp_asr}

We evaluate timestamp-aware ASR using \textbf{Accumulated Average Shift (AAS)}, following the metric used in Qwen3-ASR~\cite{shi2026qwen3asr}. AAS measures the average absolute time shift between predicted timestamps and reference timestamps over all evaluated timestamp slots:
\[
\mathrm{AAS}
=
\frac{1}{N}
\sum_{i=1}^{N}
\left|
\hat{t}_i - t_i
\right|,
\]
where \(N\) is the total number of timestamp slots, \(\hat{t}_i\) is the predicted timestamp for the \(i\)-th slot, and \(t_i\) is the corresponding reference timestamp. Lower AAS indicates more accurate temporal alignment. In our evaluation, AAS is reported in milliseconds.

We construct the timestamp ASR test sets from the official test sets of \textbf{AISHELL-1} and \textbf{LibriSpeech}. Since these datasets provide high-quality transcriptions but not word-level timestamp annotations in the required format, we first apply CTC alignment to the audio--transcript pairs to obtain reference timestamp labels. The resulting aligned annotations are then used as the reference timestamps for evaluating model outputs. This allows us to measure not only whether the model transcribes the speech correctly, but also whether it places the recognized content at the correct time positions.

\begin{table}[htbp]
\centering
\caption{\textbf{Timestamp ASR results measured by AAS.} Lower is better.}
\label{tab:timestamp_asr}
\small
\setlength{\tabcolsep}{8pt}
\renewcommand{\arraystretch}{1.08}
\begin{tabular}{@{}>{\raggedright\arraybackslash}p{4.0cm}cc@{}}
\toprule
\multicolumn{1}{c}{\textbf{Model}} & \textbf{AISHELL-1 (zh)} & \textbf{LibriSpeech (en)} \\
\midrule
Audio-Flamingo-Next & -- & 211.15 \\
Qwen3-Omni-Instruct & 833.66 & 646.95 \\
Gemini-3.1-Pro & 708.24 & 871.19 \\
\textbf{MOSS-Audio-4B-Instruct} & 76.96 & 358.13 \\
\textbf{MOSS-Audio-8B-Instruct} & \textbf{35.77} & \textbf{131.61} \\
\bottomrule
\end{tabular}
\end{table}

As shown in Table~\ref{tab:timestamp_asr}, \textbf{MOSS-Audio-8B-Instruct} achieves the strongest timestamp ASR performance among the compared models. The results show that MOSS-Audio can produce accurate time-aligned transcriptions on both Chinese and English speech. Compared with general omni-model baselines, MOSS-Audio obtains substantially lower AAS, indicating that its time-aware pretraining and timestamp ASR supervision effectively improve temporal alignment rather than only improving transcription fluency.

\section{Related Work}

\textbf{Unified speech-text modeling.}
Early speech-language models established that speech and text can be modeled within a shared sequence-to-sequence or language-modeling framework instead of being handled by isolated task-specific systems. 
SpeechT5~\cite{ao2021speecht5} unifies speech and text processing with a shared encoder-decoder backbone and modality-specific pre/post-nets, supporting a broad set of spoken language processing tasks. 
Unified Speech-Text Pre-training~\cite{tang2022unified} further studies joint speech-text pre-training for speech recognition and speech translation by combining self-supervised speech learning, text modeling, and supervised cross-modal objectives. 
More recent systems extend this line toward spoken interaction and conversational interfaces. 
SpeechGPT~\cite{zhang2023speechgpt} discretizes speech into token sequences and incorporates them into a large language model for cross-modal instruction following and speech interaction, while SPIRIT-LM~\cite{nguyen2025spirit} studies interleaved spoken and written language modeling by continuously training a text language model on text, speech, and aligned speech-text sequences. 
Moshi~\cite{defossez2024moshi} models full-duplex spoken dialogue through parallel speech streams and neural audio codec tokens, reducing the dependence on cascaded ASR--LLM--TTS pipelines. 
Mini-Omni~\cite{xie2024miniomni}, GLM-4-Voice~\cite{zeng2024glm}, Baichuan-Audio~\cite{li2025baichuan}, and Step-Audio~\cite{huang2025step} further explore real-time speech interaction, controllable speech generation, and unified speech understanding-generation systems. 
These works demonstrate the feasibility of bringing speech into language-model-style generation and interaction, but their primary focus is speech-text conversion, spoken dialogue, or speech generation rather than unified understanding of speech, music, and general audio.

\textbf{Large audio-language models.}
A second line of work extends language models from speech-centric processing to broader audio understanding. 
LTU-AS~\cite{gong2023jointaudio} combines a speech/audio perception module with an LLM to jointly understand spoken content, paralinguistic information, and non-speech audio events. 
SALMONN~\cite{tang2024salmonn} integrates speech and audio encoders with a text LLM to support speech, audio event, and music understanding within one model. 
Qwen-Audio~\cite{chu2023qwenaudio} scales audio-language pre-training over many tasks and audio types, and Qwen2-Audio~\cite{chu2024qwen2} improves instruction following through natural-language prompting and larger-scale training. 
Audio Flamingo~\cite{kong2024audioflamingo}, Audio Flamingo 2~\cite{ghosh2025audioflamingo2}, and Audio Flamingo Next~\cite{ghosh2026audioflamingonext} emphasize audio understanding, few-shot adaptation, dialogue, long-audio processing, and audio reasoning. 
GAMA~\cite{ghosh2024gama} similarly targets advanced audio understanding and complex reasoning by combining LLMs with richer audio representations and audio-language instruction data. 
Broader omni-modal systems such as Qwen2.5-Omni~\cite{qwen2.5omni} and Qwen3-Omni~\cite{qwen3omni} further integrate text, image, audio, and video perception with text and speech generation, using architectures such as Thinker--Talker and modality-specific streaming designs. 
These systems show that audio-language modeling is moving from recognition toward open-ended audio reasoning and interactive agents. 
MOSS-Audio follows this direction, but is designed specifically as an understanding-centric model family for speech, environmental sound, and music, with explicit support for captioning, timestamped transcription, time-aware question answering, and reasoning in one autoregressive text-generation framework. 
Speech generation or tool-mediated response generation can be built downstream on top of this perceptual and reasoning foundation.

\textbf{Audio representation learning.}
The quality of audio-language modeling depends strongly on the audio representation exposed to the language model. 
Self-supervised speech encoders such as HuBERT~\cite{hsu2021hubert} and WavLM~\cite{chen2022wavlm} learn robust representations for speech recognition and full-stack speech processing. 
For general audio, BEATs~\cite{chen2023beats} learns bidirectional audio representations with acoustic tokenizers, while CLAP~\cite{elizalde2023clap} and LAION-CLAP~\cite{wu2023laionclap} align audio and natural language through contrastive pre-training. 
In parallel, neural audio codecs and speech tokenizers, including EnCodec~\cite{defossez2022high}, improved RVQGAN codecs~\cite{kumar2023high}, and SpeechTokenizer~\cite{zhang2023speechtokenizer}, provide discrete or compressed representations that are useful for speech generation and speech language modeling. 
Recent analyses of neural audio codecs~\cite{ye2025codec} indicate that representations optimized for reconstruction, contrastive retrieval, or speech synthesis do not necessarily preserve all levels of acoustic evidence needed for fine-grained understanding, temporal localization, and reasoning. 
MOSS-Audio therefore uses a dedicated audio encoder and injects multi-level encoder features into the language model. 
This design follows a broader trend in multimodal language models that mitigates the bottleneck of exposing the decoder only to a single final encoder representation. 
DeepStack~\cite{meng2024deepstack} injects additional visual token features into intermediate language-model layers, and Qwen3-VL~\cite{bai2025qwen3vl} further adopts DeepStack integration to leverage multi-level ViT features for stronger vision-language alignment. 
MOSS-Audio adapts this principle to audio by routing multi-level encoder states into the language model, preserving both low-level acoustic cues and high-level semantic evidence for downstream audio reasoning.

\textbf{Temporal grounding and time-aware audio modeling.}
Temporal grounding has become an increasingly important capability for audio-language models. 
Beyond recognizing what is present in an audio clip, a model should also determine when events occur, how speaker turns evolve over time, and which acoustic evidence supports a time-sensitive answer. 
Timestamped transcription has long been a practical target in speech recognition systems such as Whisper~\cite{radford2023robust}, while recent end-to-end speaker-attributed transcription systems such as MOSS-Transcribe-Diarize~\cite{yu2026mosstranscribediarize} further emphasize the need to jointly model lexical content, speaker identity, and timestamps over long recordings. 
Recent audio-language modeling work suggests that explicit time representations can make temporal grounding easier than relying only on latent positional information. 
TimeAudio~\cite{wang2025timeaudio} introduces temporal markers and absolute time-aware encoding to connect audio semantics with precise temporal perception, while SpotSound~\cite{sun2026spotsound} interleaves textual timestamp tokens with audio embeddings to support event-boundary localization for open-vocabulary audio queries. 
Related ideas have also appeared in video-language models: TimeMarker~\cite{chen2024timemarker} uses temporal separator tokens to encode absolute frame positions, and Qwen3-VL~\cite{bai2025qwen3vl} adopts explicit textual timestamp alignment for more precise video temporal grounding. 
MOSS-Audio follows this line of work by inserting explicit time markers into the audio representation sequence during pretraining, enabling the model to learn not only what happens in the audio, but also when it happens, and supporting timestamp-aware ASR, event localization, and time-based audio question answering.

\section{Conclusion}

This report presented \textbf{MOSS-Audio}, a unified audio-language model family for speech understanding, environmental sound understanding, music understanding, audio captioning, time-aware question answering, and complex reasoning. MOSS-Audio combines a dedicated audio encoder, DeepStack cross-layer feature injection, explicit time-aware representation, a branched annotation pipeline over speech, music, and general audio, and a staged training recipe that separates instruction following from deeper reasoning behavior.

The resulting family already shows a strong and distinctive empirical profile. \textbf{MOSS-Audio-8B-Thinking} achieves the strongest results on the broad general audio understanding suite, while \textbf{MOSS-Audio-8B-Instruct} performs best on speech captioning, ASR, and timestamp ASR. These results suggest that understanding-centric audio modeling can support both open-domain acoustic comprehension and precise speech-oriented tasks within a single model family.

More broadly, MOSS-Audio shows that a single model can cover descriptive understanding, transcription, temporal grounding, and harder reasoning over heterogeneous audio without breaking into disconnected specialist systems. This makes it a promising foundation for future \textbf{voice agents}: rather than serving only as a transcription or captioning module, MOSS-Audio can act as the audio-understanding core that perceives user intent, acoustic context, temporal events, and reasoning-relevant cues, and can be connected with tools such as dialogue systems, retrieval, action execution, and speech generation to build more capable real-time interactive agents.

\clearpage
\section*{Contributors}

\noindent\textbf{Core Contributors}: \\
Chen Yang$^{*}$, Chufan Yu, Hanfu Chen, Jie Zhu, Jingqi Chen, Ke Chen, Wenxuan Wang, Yang Wang, Yaozhou Jiang, Yi Jiang, Zhengyuan Lin,  Ziqi Chen, Zhaoye Fei$^{*}$

\vspace{0.5em}
\noindent\textbf{Contributors}: \\
Chenghao Liu, Donghua Yu, Jun Zhan, Kang Yu, Kexin Huang, Liwei Fan, Mingshu Chen, Qinyuan Cheng, Ruixiao Li, Shimin Li, Songlin Wang, Xingjian Zhao, Yang Gao, Yitian Gong, Yiyang Zhang, Zhe Xu

\vspace{0.5em}

\noindent\textbf{Advisors}: \\
Xipeng Qiu$^{\S}$

\vspace{1em}

\noindent\textbf{Affiliations}: \\
Shanghai Innovation Institute\\
MOSI Intelligence\\
Fudan University\\


\begingroup
\renewcommand{\thefootnote}{}
\makeatletter
\renewcommand{\@makefntext}[1]{\noindent #1}
\makeatother
\footnotetext{
\textsuperscript{*}Project Lead.
\textsuperscript{\S}Corresponding Author: \texttt{xpqiu@fudan.edu.cn}.\\
All Contributors are sorted alphabetically by first name.
}
\endgroup


\clearpage

\bibliographystyle{unsrtnat}
\bibliography{main}
\clearpage
\appendix
\section{Additional Details}
\subsection{Evaluation Prompts}
\label{sec:evaluation_prompts}

\newtcolorbox{mossbox}[1]{
    enhanced,
    colback=white,
    colframe=MossBlue,
    colbacktitle=MossBlue,
    coltitle=black,
    title style={left color=MossCyan, right color=MossBlue},
    title=\textbf{#1},
    fontupper=\ttfamily,
    boxrule=0.8pt,
    left=2mm, right=2mm, top=2mm, bottom=2mm
}

\begin{mossbox}{Shared Audio-Text Evaluation Template}
[system]\\
You are a helpful assistant.

[user]\\
<audio>\\
\{task-specific instruction or question\}

[assistant]\\
\{model response to be evaluated\}
\end{mossbox}

\begin{mossbox}{Automatic Speech Recognition (ASR)}
[system]\\
You are a helpful assistant.

[user]\\
<audio>\\
Transcribe the audio into text.

[assistant]\\
Return only the transcription of the speech content. Do not include explanations, headings, or any additional comments.
\end{mossbox}

\begin{mossbox}{Speech Captioning}
[system]\\
You are a helpful assistant.

[user]\\
<audio>\\
用 JSON 格式描述音频中的人声特征，包含 gender、age、pitch、speed、volume、clarity、fluency、accent、texture、emotion、tone、personality 和 summary 字段，每个字段填写对应特征的描述。

[assistant]\\
Return a valid JSON object with the required fields: gender, age, pitch, speed, volume, clarity, fluency, accent, texture, emotion, tone, personality, and summary.
\end{mossbox}

\begin{mossbox}{Sentence-Level Timestamp ASR}
[system]\\
You are a helpful assistant.

[user]\\
<audio>\\
Transcribe the audio into sentence-level timestamps using the format [xx.xx]text[yy.yy], where xx.xx and yy.yy are millisecond-precision timestamps rounded to two decimal places. Concatenate consecutive segments without breaks, for example [0.00]hello there[2.01][2.01]the weather is nice today[4.32]. Exclude any additional content such as explanations, headings, or annotations.

[assistant]\\
Return only the timestamped transcript in the required format.
\end{mossbox}

\begin{mossbox}{Word-Level Timestamp ASR}
[system]\\
You are a helpful assistant.

[user]\\
<audio>\\
可以将这段音频逐字按时间戳转写出来吗？每个字都要写成 [xx.xx]文本[yy.yy] 的形式，时间戳用两位小数的毫秒数，英文方括号括起来，片段之间直接连起来，不要加其他内容。

[assistant]\\
Return only the word-level timestamped transcript. Do not add explanations, headings, or line breaks.
\end{mossbox}

\begin{mossbox}{Open-Ended Audio Question Answering}
[system]\\
You are a helpful assistant.

[user]\\
<audio>\\
\{question\}

[assistant]\\
Return the answer to the question based on the audio content.
\end{mossbox}

\begin{mossbox}{Multiple-Choice Audio Question Answering}
[system]\\
You are a helpful assistant.

[user]\\
<audio>\\
Choose the most suitable answer from options A, B, C, and D to respond to the question in the next line. You should only choose A, B, C, or D. Do not provide any additional explanations or content.\\
Question: \{question with options\}

[assistant]\\
Return only one option letter: A, B, C, or D.
\end{mossbox}

\subsection{Timestamp Serialization Examples}
\label{sec:timestamp_examples} 
To illustrate the data formats used in our experiments, we provide examples of word-level and sentence-level timestamped transcripts below.

\begin{mossbox}{Word-level Timestamp Example}
[0.02]海[0.12][0.16]风[0.36][0.36]迎[0.56][0.62]面，[0.8][0.92]列[1.06][1.22]车[1.36][1.54]奔[1.72][1.8]跑[2.06][2.14]在[2.36][2.44]这[2.62][3.02]条[3.3][3.34]海[3.6][3.7]岸[3.82][3.94]线，[4.36][4.84]想[5.12][5.16]与[5.22][5.4]家[5.52][5.62]人[6][6]一[6.28][6.28]起[6.44][6.62]度[6.74][6.92]过[7.04][7.5]这[7.62][7.72]美[8.04][8.12]丽[8.22][8.36]夏[8.52][8.72]天。[9.14][9.6]就[9.74][9.74]在[9.9][9.9]列[10.08][10.2]车[10.32][10.48]冲[10.72][10.76]出[10.9][11.09]隧[11.32][11.4]道，[11.62][11.66]迎[11.95][11.98]来[12.46][12.61]阳[12.85][12.93]光[13.16][13.24]时[13.39][13.53]刻，[13.65][14.47]草[14.69][14.77]帽[14.99][15.05]调[15.25][15.25]皮[15.37][15.63]单[15.75][15.95]飞[16.15][16.23]向[16.51][16.57]了[16.71][17.15]碧[17.29][17.45]空[17.63][17.63]的[17.75][18.05]蓝[18.29][18.35]天。[18.71][20.45]翻[20.65][20.73]开[21.01][21.23]儿[21.25][21.35]时[21.49][21.65]那[21.77][21.93]些[22.07][22.25]绘[22.39][22.43]画[22.57][22.79]图[22.91][23.13]卷。[23.65]
\end{mossbox}

\begin{mossbox}{Sentence-level Timestamp Example}
[0.02]海风迎面，列车奔跑在这条海岸线，想与家人一起度过这美丽夏天。[9.14][9.6]就在列车冲出隧道，迎来阳光时刻，草帽调皮的飞向了碧空的蓝天。[18.71][20.45]翻开儿时那些绘画图卷。[23.65]
\end{mossbox}

\subsection{Complete ASR Evaluation Results}
\label{sec:complete_asr_results}

We provide the complete dataset-level ASR results in Table~\ref{tab:asr_detail}. Results are reported as CER (\%), where lower values indicate better recognition accuracy. For grouped datasets, tuple-style entries are generally retained in a single row to preserve the original benchmark structure; AISHELL-6A is split into two rows for readability, with each tuple position corresponding to the dataset order shown in parentheses. The best result in each row, or each tuple position for grouped entries, is highlighted in bold.

\providecommand{\mossaudiomodelheader}[1]{\shortstack[c]{#1}}
\begin{landscape}
\begin{table*}[htbp]
\centering
\caption{\textbf{Detailed ASR results.} Rows correspond to evaluation datasets or dataset groups and columns correspond to models. Lower CER is better. Tuple-style entries are generally kept in a single row to preserve the original benchmark structure; AISHELL-6A is split into two rows for readability.}
\label{tab:asr_detail}
\scriptsize
\setlength{\tabcolsep}{0.75pt}
\renewcommand{\arraystretch}{1.06}
\resizebox{\linewidth}{!}{%
\begin{tabular}{@{}lccccccccccc@{}}
\toprule
\textbf{Dataset} & \mossaudiomodelheader{\textbf{Paraformer}\\\textbf{-Large}} & \mossaudiomodelheader{\textbf{GLM-ASR}\\\textbf{-Nano}} & \mossaudiomodelheader{\textbf{Fun-ASR}\\\textbf{-Nano}} & \mossaudiomodelheader{\textbf{SenseVoice}\\\textbf{-Small}} & \mossaudiomodelheader{\textbf{Kimi-Audio}\\\textbf{-7B-Instruct}} & \mossaudiomodelheader{\textbf{Audio Flamingo}\\\textbf{Next}} & \mossaudiomodelheader{\textbf{Qwen2.5}\\\textbf{-Omni-3B}} & \mossaudiomodelheader{\textbf{Qwen2.5}\\\textbf{-Omni-7B}} & \mossaudiomodelheader{\textbf{Qwen3-Omni}\\\textbf{-Instruct}} & \mossaudiomodelheader{\textbf{MOSS-Audio}\\\textbf{-4B-Instruct}} & \mossaudiomodelheader{\textbf{MOSS-Audio}\\\textbf{-8B-Instruct}} \\
\midrule
AISHELL-1 & 1.98 & 2.89 & 2.16 & 3.23 & \textbf{0.79} & 6.85 & 1.51 & 1.16 & 0.95 & 2.26 & 1.82 \\
AISHELL-2 (Android|iOS|Mic) & 3.28|3.21|3.00 & 3.75|3.73|3.78 & 3.04|2.99|3.07 & 4.16|4.02|3.96 & 2.91|3.03|2.88 & 7.22|8.12|8.62 & 3.10|2.94|2.93 & 2.88|2.77|2.73 & \textbf{2.70|2.72|2.57} & 3.22|3.20|3.33 & 2.97|2.95|2.91 \\
THCHS-30 & 4.07 & 4.23 & 3.65 & 5.26 & \textbf{1.39} & 12.41 & 3.32 & 3.06 & 2.21 & 3.53 & 2.82 \\
MAGICDATA-READ & 4.67 & 5.02 & 3.46 & 4.93 & \textbf{2.15} & 6.24 & 3.56 & 3.16 & 2.47 & 3.72 & 3.20 \\
AISHELL6-Whisper (normal|whisper) & 1.11|8.92 & 0.83|9.06 & 0.81|6.76 & 1.25|9.88 & 0.69|4.63 & 6.48|19.20 & 0.82|7.82 & 0.71|6.57 & \textbf{0.59|3.22} & 0.73|5.86 & 0.69|4.80 \\
AliMeeting (Test\_Ali\_far|Test\_Ali\_near) & \textbf{25.64}|9.27 & 40.27|14.76 & 27.21|9.55 & 37.01|16.31 & 28.22|13.82 & 60.68|25.12 & 32.14|12.16 & 32.03|18.73 & 25.72|\textbf{8.44} & 27.27|9.68 & 36.82|11.25 \\
AISHELL-4 & 20.33 & 28.02 & 19.82 & 24.06 & 20.61 & 62.13 & 22.91 & 21.01 & \textbf{18.15} & 20.33 & 24.36 \\
SeniorTalk (sentence) & 17.31 & 20.33 & 16.96 & 21.07 & 19.70 & 39.44 & 17.38 & 19.96 & \textbf{14.13} & 16.93 & 17.42 \\
ChildMandarin & 12.60 & 14.06 & 12.94 & 14.18 & 13.79 & 23.06 & 12.96 & 12.29 & \textbf{8.79} & 13.25 & 13.10 \\
AISHELL-6A (mild | moderate) & 6.98|9.30 & 8.74|12.11 & 6.60|\textbf{8.81} & 7.62|9.85 & 7.00|9.34 & 24.13|29.40 & 6.87|10.55 & 7.27|10.94 & 6.20|8.88 & 6.36|9.77 & \textbf{5.84}|8.94 \\
AISHELL-6A (severe | StutteringSpeech) & 13.34|10.74 & 14.38|12.29 & 12.98|10.30 & 14.39|11.47 & 12.56|10.75 & 35.69|21.90 & 14.57|11.33 & 12.92|10.53 & 11.59|10.25 & 12.68|10.28 & \textbf{11.52}|\textbf{9.72} \\
AISHELL\_6B (LRDWWS|Uncontrol) & 47.59|45.08 & 50.34|49.09 & 47.42|45.84 & 52.92|47.97 & 44.44|42.57 & 56.17|49.52 & 54.54|50.03 & 51.99|49.45 & 45.80|41.65 & 43.35|44.25 & \textbf{39.76|39.27} \\
WenetSpeech (test-meeting) & 7.88 & 9.70 & 7.39 & 8.35 & 7.15 & 54.46 & 9.04 & 8.43 & \textbf{6.64} & 8.17 & 7.86 \\
Fleurs (cmn\_hans\_cn) & 6.40 & 4.94 & \textbf{4.76} & 6.75 & 5.10 & 12.86 & 5.45 & 5.13 & 4.84 & 8.13 & 7.52 \\
CS-Dialogue & 10.64 & 11.06 & 10.47 & 12.81 & 14.56 & 30.52 & 10.78 & 14.02 & 12.94 & 9.14 & \textbf{9.07} \\
TALCS & 10.77 & 11.07 & \textbf{8.09} & 10.52 & 12.74 & 40.42 & 10.94 & 10.46 & 8.33 & 8.37 & 8.22 \\
ASCEND & 16.55 & 13.50 & 15.13 & 18.38 & 21.83 & 32.63 & 13.25 & 14.42 & \textbf{12.64} & 12.83 & 13.26 \\
KeSpeech & 11.48 & 9.72 & 7.43 & 10.45 & \textbf{5.51} & 16.97 & 7.67 & 6.40 & 5.87 & 14.65 & 9.18 \\
WSYue-ASR-eval (short) & 75.42 & 35.07 & 8.17 & \textbf{7.34} & 53.17 & 34.74 & 60.06 & 57.43 & 25.39 & 9.04 & 8.33 \\
MIR-1K & 57.70 & 95.87 & 35.85 & 39.51 & 38.35 & 40.06 & 45.00 & 42.62 & 30.81 & 18.47 & \textbf{17.24} \\
openc-pop & 6.98 & 8.03 & 2.84 & 8.07 & 5.17 & 11.54 & 3.47 & 2.75 & \textbf{1.21} & 3.10 & 2.39 \\
MNV\_17 & 4.95 & 4.65 & 4.76 & 4.92 & 4.68 & 8.25 & 5.54 & 4.56 & 4.73 & \textbf{4.01} & 4.31 \\
\bottomrule
\end{tabular}%
}
\end{table*}
\end{landscape}

\subsection{Ablation Study}
In this section, we present comprehensive ablation studies to validate the key architectural designs and pre-training strategies of MOSS-Audio. Specifically, we investigate the model along three critical dimensions: (1) the general audio representation capability of the MOSS Audio Encoder across diverse domains, (2) its fundamental speech recognition (ASR) ceiling under strictly controlled settings, and (3) the effectiveness of the DeepStack feature injection mechanism in preserving non-speech acoustic cues. Together, these experiments provide empirical evidence for the design choices that enable MOSS-Audio's holistic understanding capabilities.

\subsubsection{Audio Representation Capability}

To rigorously assess the representation quality of our pre-trained audio backbone, we conduct comparative experiments using the XARES-LLM framework\cite{dinkel2026interspeech2026audioencoder}, a holistic evaluation suite which trains a typical LALM(Large Audio Language Model) using the audio encoder provided by the user.  We evaluate the MOSS Audio Encoder against the encoder in \texttt{whisper-large-v3} and the AuT encoder in \texttt{Qwen3-Omni-30B-A3B-Instruct}. 

As shown in Table \ref{tab:encoder_ablation}, the evaluation is structured into two main tracks. Task 1 assesses general audio understanding across 15 diverse benchmarks (e.g., environmental sounds, music genre, speaker verification). Here, the MOSS Audio Encoder maintains highly competitive performance, significantly outperforming \texttt{whisper-large-v3} overall and achieving the best results on complex datasets such as ASVspoof, ESC-50, and FSD50k. 

Furthermore, on Task 2, which emphasizes downstream generative capabilities including Automatic Speech Recognition (ASR) and Audio Captioning, our encoder demonstrates clear superiority. It achieves an overall state-of-the-art score of 0.673, consistently outperforming both baselines. The gains are particularly pronounced in speech transcription (AISHELL-1, LibriSpeech) and natural language audio description (Clotho). This strongly affirms that our pre-training strategy effectively integrates rich acoustic information, yielding a versatile backbone capable of supporting both fine-grained perception and high-quality generation.
\begin{table*}[t]
\centering
\caption{\textbf{Ablation on Audio Encoder Capability.} We evaluate the representations of different audio encoders using the XARES-LLM framework \cite{dinkel2026interspeech2026audioencoder}. Task 1 focuses on audio classification and understanding, while Task 2 evaluates generative capabilities including ASR (1-WER/CER) and Audio Captioning (FENSE/DATE). Best results are highlighted in bold.}
\label{tab:encoder_ablation}

\resizebox{\textwidth}{!}{
\begin{tabular}[htbp]{lccccccccccccccc|c}
\toprule
\multirow{2}{*}{Encoder} & \multicolumn{16}{c}{Task 1: Audio Understanding Benchmark} \\
\cmidrule{2-17}
& ASV & CRE & ESC & FSC & FMA & F50k & FKag & GTZ & LbC & NSy & SPC & U8k & Voc & VC1 & VL & AVG$_1$ \\
\midrule
Whisper-large-v3\cite{radford2023robust} & 0.958 & 0.722 & 0.715 & 0.946 & 0.637 & 0.123 & 0.670 & 0.859 & 0.460 & 0.698 & 0.723 & 0.785 & 0.913 & 0.962 & \textbf{0.973} & 0.743 \\
AuT\cite{qwen3omni} & 0.971 & \textbf{0.777} & 0.770 & \textbf{0.996} & \textbf{0.660} & 0.158 & \textbf{0.770} & \textbf{0.869} & \textbf{0.736} & \textbf{0.767} & \textbf{0.974} & \textbf{0.843} & \textbf{0.934} & 0.955 & 0.953 & \textbf{0.809} \\
MOSS Audio Encoder & \textbf{0.986} & 0.678 & \textbf{0.785} & \textbf{0.996} & 0.623 & \textbf{0.164} & 0.733 & 0.788 & 0.488 & 0.712 & 0.828 & 0.822 & 0.928 & \textbf{0.974} & 0.902 & 0.760 \\
\bottomrule
\multicolumn{17}{l}{\scriptsize \textit{Abbreviations for Task 1} -- \textbf{ASV}: ASVspoof2015; \textbf{CRE}: Crema-D; \textbf{ESC}: ESC-50; \textbf{FSC}: FluentSpeechCommands; \textbf{FMA}: FreeMusicArchive;} \\
\multicolumn{17}{l}{\scriptsize \textbf{F50k}: FSD50k; \textbf{FKag}: FSDKaggle2018; \textbf{GTZ}: GTZAN; \textbf{LbC}: LibriCount; \textbf{NSy}: NSynth; \textbf{SPC}: SpeechCommandsV1; \textbf{U8k}: UrbanSound8K;} \\
\multicolumn{17}{l}{\scriptsize \textbf{Voc}: VocalSound; \textbf{VC1}: VoxCeleb1; \textbf{VL}: VoxLingua33.} \\
\end{tabular}
}

\vspace{0.3cm} 

\resizebox{0.7\textwidth}{!}{
\begin{tabular}[htbp]{lccccc|c}
\toprule
\multirow{2}{*}{Encoder} & \multicolumn{6}{c}{Task 2: Speech \& Generation Benchmark} \\
\cmidrule{2-7}
& AISHELL & Clotho & LibriSpeech & MECAT & SongDesc & AVG$_2$ \\
\midrule
Whisper-large-v3\cite{radford2023robust} & 0.467 & 0.402 & 0.458 & 0.652 & 0.480 & 0.492 \\
AuT\cite{qwen3omni} & 0.849 & 0.399 & 0.899 & 0.654 & \textbf{0.505} & 0.661 \\
MOSS Audio Encoder & \textbf{0.888} & \textbf{0.411} & \textbf{0.918} & \textbf{0.657} & 0.491 & \textbf{0.673} \\
\bottomrule
\end{tabular}
}
\end{table*}

\subsubsection{In-Depth ASR Capability Analysis}

To rigorously probe the fundamental speech recognition ceiling, we conduct an in-depth, strictly controlled ASR evaluation. We compare the MOSS Audio Encoder against the baseline Audio Transformer \cite{qwen3omni} (AuT) in \texttt{Qwen3-Omni-30B-A3B-Instruct}. Each encoder is integrated with a \texttt{Qwen3-1.7B} language model and pretrained on the same pretraining data as MOSS-Audio for 100k steps.

The models are evaluated across an extensive suite of 38 diverse speech test sets, encompassing standard read speech, noisy environments, multi-speaker meetings, dialectal accents, and highly challenging atypical speech (e.g., whispered, stammering, and singing). 

As detailed in Table \ref{tab:asr_indepth}, the MOSS Audio Encoder demonstrates a profound and consistent advantage. It reduces the average CER or WER across all 38 datasets from 17.61\% to 16.31\%. Notably, the performance gains are most dramatic in extreme acoustic scenarios where standard models typically struggle. For instance, on whispered speech (\texttt{AISHELL6-Whisper}), the error rate drops sharply from 12.45\% to 7.87\%; on singing voice (\texttt{Opencpop}), it decreases from 4.91\% to 3.43\%; and on severe stammering (\texttt{AISHELL-6A/severe}), it improves from 17.31\% to 14.99\%. These results empirically validate that the MOSS Audio Encoder captures phonetic and acoustic nuances significantly better than the vanilla AuT, elevating the overall generative ceiling of the connected LLM.

\begin{table*}[t]
\centering
\caption{\textbf{In-Depth ASR Performance Comparison.} Both AuT and MOSS Audio Encoder were paired with \texttt{Qwen3-1.7B} and pretrained for 100k steps. Results are reported as CER or WER (\%), where lower is better. The MOSS Audio Encoder yields lower error rates on 36 out of 38 datasets.}
\label{tab:asr_indepth}
\resizebox{\textwidth}{!}{
\begin{tabular}{lcc c lcc}
\toprule
\multirow{2}{*}{\textbf{Dataset}} & \multicolumn{2}{c}{\textbf{CER or WER (\%) $\downarrow$}} & & \multirow{2}{*}{\textbf{Dataset}} & \multicolumn{2}{c}{\textbf{CER or WER (\%) $\downarrow$}} \\
\cmidrule(lr){2-3} \cmidrule(lr){6-7}
& \textbf{AuT\cite{qwen3omni}} & \textbf{MOSS Audio Encoder} & & & \textbf{AuT \cite{qwen3omni}} & \textbf{MOSS Audio Encoder} \\
\midrule
AISHELL-1 & 3.28 & \textbf{2.58} & & ChildMandarin & \textbf{14.54} & 15.14 \\
AISHELL-2/Android & 4.19 & \textbf{3.50} & & CommonVoice & 10.00 & \textbf{9.25} \\
AISHELL-2/Mic & 4.14 & \textbf{3.58} & & KeSpeech & \textbf{15.88} & 16.16 \\
AISHELL-2/iOS & 3.94 & \textbf{3.46} & & MAGICDATA-READ & 4.93 & \textbf{4.60} \\
AISHELL-4 & 29.38 & \textbf{27.88} & & MIR-1K & 23.65 & \textbf{22.90} \\
AISHELL-5/Eval1 & 38.55 & \textbf{38.37} & & MMedFD/Agent & 45.05 & \textbf{44.98} \\
AISHELL-5/Eval2 & 43.46 & \textbf{40.93} & & MMedFD/User & 4.98 & \textbf{4.76} \\
AISHELL-6A/Stammer/mild & 9.63 & \textbf{8.13} & & MNV\_17 & 4.97 & \textbf{4.58} \\
AISHELL-6A/Stammer/moderate & 14.01 & \textbf{12.01} & & MagicData-RAMC & 16.12 & \textbf{14.08} \\
AISHELL-6A/Stammer/severe & 17.31 & \textbf{14.99} & & Opencpop & 4.91 & \textbf{3.43} \\
AISHELL-6A/StutteringSpeech & 14.28 & \textbf{12.95} & & SeniorTalk/dialogue & 27.62 & \textbf{26.54} \\
AISHELL-6B/LRDWWS & 59.66 & \textbf{56.08} & & SeniorTalk/sentence & 21.56 & \textbf{20.02} \\
AISHELL-6B/MDSC/Uncontrol & 57.84 & \textbf{54.33} & & TALCS & 10.92 & \textbf{8.69} \\
AISHELL6-Whisper/normal & 1.19 & \textbf{0.87} & & THCHS-30 & 3.91 & \textbf{3.50} \\
AISHELL6-Whisper/whisper & 12.45 & \textbf{7.87} & & WSYue-ASR-eval/short & 11.25 & \textbf{10.74} \\
ASCEND & 16.43 & \textbf{14.50} & & Wenet\_Speech/test\_meeting & 9.55 & \textbf{8.45} \\
AliMeeting/Test\_Ali\_far & 43.38 & \textbf{41.40} & & Wenet\_Speech/test\_net & 11.43 & \textbf{9.42} \\
AliMeeting/Test\_Ali\_near & 13.11 & \textbf{11.41} & & WildElder & 22.86 & \textbf{19.69} \\
CS-Dialogue/short\_wav & 10.70 & \textbf{10.14} & & fleurs/cmn\_hans\_cn & 8.10 & \textbf{7.75} \\
\midrule
\multicolumn{5}{l}{\textbf{Overall Average across all 38 datasets}} & 17.61 & \textbf{16.31} \\
\bottomrule
\end{tabular}
}
\end{table*}

\subsubsection{Ablation on DeepStack}

To verify the effectiveness of the DeepStack mechanism, we conduct an ablation study in a controlled setting. We pair the MOSS Audio Encoder with a lightweight language model (\texttt{Qwen3-0.6B-base}). The baseline model utilizes only the final layer of the audio encoder, while our proposed method injects intermediate layer features via DeepStack. Both models undergo an identical two-stage training pipeline: an initial alignment phase using ASR data, followed by fine-tuning on the MECAT-Caption \cite{mecat2025} dataset. 

The models are evaluated on the MECAT-Caption test set using the DATE \cite{mecat2025} metric (higher is better) across various fine-grained acoustic scenarios. As demonstrated in Table \ref{tab:deepstack_ablation}, the DeepStack mechanism yields an overall performance improvement. Interestingly, while there is a slight degradation in speech-dominated scenarios (Pure and Mixed Speech), the model achieves consistent and notable gains across all non-speech categories, including music, pure sound, and environmental acoustics. 

This trade-off strongly aligns with our architectural intuition. The top layer of the audio encoder is highly specialized for speech semantics due to the ASR alignment phase. By explicitly injecting intermediate features through DeepStack, we successfully rescue the low-level acoustic, timbral, and environmental cues that would otherwise be overshadowed, effectively boosting the model's holistic audio understanding capability without requiring additional parameters in the audio backbone.
\begin{table*}[t]
\centering
\caption{\textbf{Ablation on DeepStack Feature Injection.} Evaluated on the MECAT-Caption\cite{mecat2025} test set using the DATE metric. The models (MOSS Audio Encoder + Qwen3-0.6B-base) are compared between using only the final encoder layer (Baseline) and injecting intermediate layers (DeepStack). Best results are highlighted in bold.}
\label{tab:deepstack_ablation}
\resizebox{\textwidth}{!}{
\begin{tabular}{lcccccccccc}
\toprule
\multirow{2}{*}{Method} & \multicolumn{2}{c}{Content Length} & \multicolumn{2}{c}{Speech} & \multicolumn{2}{c}{Music} & \multicolumn{3}{c}{Acoustic \& Environment} & \multirow{2}{*}{Overall} \\
\cmidrule(lr){2-3} \cmidrule(lr){4-5} \cmidrule(lr){6-7} \cmidrule(lr){8-10}
& Long & Short & Pure & Mixed & Pure & Mixed & Pure Sound & Mixed Sound & Env. & \\
\midrule
Baseline & 0.6961 & 0.6692 & \textbf{0.4861} & \textbf{0.4922} & 0.4705 & 0.2800 & 0.4015 & 0.1936 & 0.1586 & 0.4823 \\
DeepStack & \textbf{0.6969} & \textbf{0.6708} & 0.4791 & 0.4838 & \textbf{0.4927} & \textbf{0.2844} & \textbf{0.4255} & \textbf{0.1938} & \textbf{0.1594} & \textbf{0.4831} \\
\bottomrule
\end{tabular}
}
\end{table*}

\end{document}